\documentclass[useAMS,usedcolumn,usenatbib,usegraphicx]{mn2e}

\usepackage{epsfig,graphicx,natbib}
\usepackage{./reference/mycite}
\usepackage{amssymb}
\usepackage{amsfonts}
\usepackage{amsmath}
\usepackage{color}
\usepackage{lscape,longtable}
 
\begin{document}

\title[Stellar Masses of WiggleZ Galaxies]{The Stellar Masses of $\sim$40,000 UV Selected Galaxies from the WiggleZ Survey at 0.3$<z<$1.0: Analogues of Lyman Break Galaxies?} 
\author[]{\parbox[t]{\textwidth}{
    Manda Banerji$^{1,2}$
    Karl Glazebrook$^3$, 
    Chris Blake$^3$,
    Sarah Brough$^4$, 
    Matthew Colless$^4$, 
    Carlos Contreras$^3$,
    Warrick Couch$^3$,
    Darren J. Croton$^3$, 
    Scott Croom$^5$, 
    Tamara M. Davis$^6$, 
    Michael J.\ Drinkwater$^6$,
    Karl Forster$^7$, 
    David Gilbank$^8$,  
    Mike Gladders$^9$, 
    Ben Jelliffe$^5$, 
    Russell J.\ Jurek$^{10}$, 
    I-hui Li$^{11}$, 
    Barry Madore$^{12}$, 
    D.\ Christopher Martin$^7$,
    Kevin Pimbblet$^{13}$, 
    Gregory B.\ Poole$^{3,14}$,
    Michael Pracy$^{3,5}$, 
    Rob Sharp$^{4,15}$,
    Emily Wisnioski$^{3,16}$,
    David Woods$^{17}$, 
    Ted K.\ Wyder$^7$ and H.K.C. Yee$^{11}$} \\ \\
  $^{1}$Institute of Astronomy, University of Cambridge, Madingley Road, Cambridge, CB3 0HA, UK \\ 
  $^{2}$Department of Physics \& Astronomy, University College London, Gower Street, London WC1E 6BT, UK, \\ 
  $^3$ Centre for Astrophysics \& Supercomputing, Swinburne University of Technology, P.O. Box 218, Hawthorn, VIC 3122, Australia \\ 
  $^4$ Australian Astronomical Observatory, P.O. Box 915, North Ryde, NSW, 1670, Australia \\
  $^5$ Sydney Institute for Astronomy, School of Physics, University of Sydney, NSW 2006, Australia \\ 
  $^6$ School of Mathematics and Physics, University of Queensland, Brisbane, QLD 4072, Australia \\   
  $^7$ California Institute of Technology, MC 278-17, 1200 East California Boulevard, Pasadena, CA 91125, United States \\ 
  $^8$ South African Astronomical Observatory, PO Box 9 Observatory, 7935 South Africa \\
  $^9$ Department of Astronomy and Astrophysics, University of Chicago, 5640 South Ellis Avenue, Chicago, IL 60637, United States \\ 
  $^{10}$ Australia Telescope National Facility, CSIRO, Epping, NSW 1710, Australia \\ 
  $^{11}$ Department of Astronomy and Astrophysics, University of Toronto, 50 St.\ George Street, Toronto, ON M5S 3H4, Canada \\
  $^{12}$ Observatories of the Carnegie Institute of Washington, 813 Santa Barbara St., Pasadena, CA 91101, United States \\ 
  $^{13}$ School of Physics, Monash University, Clayton, VIC 3800, Australia \\ 
  $^{14}$ School of Physics, University of Melbourne, Parksville, VIC 3010, Australia \\
  $^{15}$ Research School of Astronomy \& Astrophysics, Australian National University, Weston Creek, ACT 2600, Australia \\ 
  $^{16}$ Max Planck Institut f\"{u}r extraterrestrische Physik, Postfach 1312, D-85748 Garching, Germany\\
  $^{17}$ Department of Physics \& Astronomy, University of British Columbia, 6224 Agricultural Road, Vancouver, BC V6T 1Z1, Canada}

\maketitle

\clearpage

\begin{abstract} 

We characterise the stellar masses and star formation rates in a sample of $\sim$40,000 spectroscopically confirmed UV luminous galaxies at 0.3$<$z$<$1.0 selected from within the WiggleZ Dark Energy Survey. In particular, we match this UV bright population to wide-field infrared surveys such as the near infrared (NIR) UKIDSS Large Area Survey (LAS) and the mid infrared \textit{WISE} All-Sky Survey. We find that $\sim$30 per cent of the UV-luminous WiggleZ galaxies, corresponding to the brightest and reddest subset, are detected at $>$5$\sigma$ in the UKIDSS-LAS at all redshifts. An even more luminous subset of 15 per cent are also detected in the \textit{WISE} 3.4 and 4.6$\mu$m bands. In addition, 22 of the WiggleZ galaxies are extremely luminous at 12 and 22$\mu$m and have colours consistent with being star formation dominated. We compute stellar masses for this very large sample of extremely blue galaxies and quantify the sensitivity of the stellar mass estimates to various assumptions made during the SED fitting. The median stellar masses are log$_{10}$(M$_*$/M$_\odot$)=9.6$\pm$0.7, 10.2$\pm$0.5 and 10.4$\pm$0.4 for the IR-undetected, UKIDSS detected and UKIDSS+\textit{WISE} detected galaxies respectively. We demonstrate that the inclusion of NIR photometry can lead to tighter constraints on the stellar masses by bringing down the upper bound on the stellar mass estimate. The mass estimates are found to be most sensitive to the inclusion of secondary bursts of star formation as well as changes in the stellar population synthesis models, both of which can lead to median discrepancies of the order of 0.3 dex in the stellar masses. We conclude that even for these extremely blue galaxies, different SED fitting codes therefore produce extremely robust stellar mass estimates. We find however, that the best-fit M/L$_K$ is significantly lower than that predicted by simple optical colour-based estimators for many of the WiggleZ galaxies. The simple colour-based estimator over-predicts M/L$_K$ by $\sim$0.4 dex on average. The effect is more pronounced for bluer galaxies with younger best-fit ages. The WiggleZ galaxies have star formation rates of 3--10 M$_\odot$yr$^{-1}$ and mostly lie at the upper end of the main sequence of star-forming galaxies at these redshifts. Their rest-frame UV luminosities and stellar masses are comparable to both local compact UV-luminous galaxies as well as Lyman break galaxies at $z\sim2-3$. The stellar masses from this paper will be made publicly available with the next WiggleZ data release.

\end{abstract}

\begin{keywords}
galaxies: stellar content, galaxies: formation, galaxies: evolution
\end{keywords}

\section{INTRODUCTION}

In recent years, large area spectroscopic surveys of both passive galaxies such as the Luminous Red Galaxy (LRG) population, as well as emission line galaxies (ELGs), have been successfully used to place accurate constraints on cosmological models (e.g. \citealt{Blake:07, Collister:08, Percival:10, Blake:11, Reid:12}). At the same time, deeper, smaller area spectroscopic surveys such as DEEP2 and VVDS as well as multi-wavelength photometric surveys such as COMBO-17 \citep{Bell:04}, CFHTLS \citep{Arnouts:07} and COSMOS \citep{Ilbert:09}, have enabled detailed studies of galaxy formation and evolution and constraints on the global properties such as the spectral energy distributions (SEDs), ages, stellar masses and star formation histories (SFHs) of galaxies out to z$\sim$2. The advent of very large area photometric surveys at multiple wavelengths now offers us the possibility of constraining the global properties of the large spectroscopic samples that have been assembled for cosmology. Although the large area surveys are by their nature very shallow, they contain huge numbers of galaxies at z$<$1, therefore enabling a statistically robust census of galaxy properties at these redshifts. 

Star-forming galaxies at the main epoch of galaxy formation at z$\sim$1--3 have been selected in many different ways. The most luminous starbursts at these epochs are often selected at long wavelengths such as the far infrared and submillimeter (e.g. \citealt{Smail:02, Ivison:02, Chapman:05, Magnelli:10, Banerji:11}) while more modest star-forming galaxies have been targeted using optical colour-cuts such as the Lyman break selection (e.g. \citealt{Pettini:01}), the BM/BX method \citep{Adelberger:04, Steidel:04} and the B$z$K technique \citep{Daddi:04}. The availability of UV data from the \textit{Galaxy Evolution Explorer (GALEX)} has also allowed comprehensive studies of UV-luminous galaxies both in the local Universe \citep{Heckman:05} and at higher redshifts \citep{Burgarella:06,Haberzettl:11}.

In this work, we study the physical properties of a population of UV-luminous emission line galaxies selected from within the WiggleZ Dark Energy Survey \citep{Drinkwater:10}. Although primarily designed as a cosmology survey targeting blue emission line galaxies at intermediate redshifts of z$\sim$0.7, the dataset contains $\sim$215,000 spectroscopically confirmed highly star-forming galaxies that form a very large statistical sample that can also be exploited for galaxy evolution studies (e.g. \citealt{Wisnioski:11}, Jurek et al. in preparation). In particular, the redshift range of the WiggleZ survey coupled with the UV selection means this sample is particularly useful for bridging the gap between analogously selected local galaxies and UV-luminous galaxies at z$\gtrsim$1. While the large sample size makes targeted multi-wavelength follow-up of the majority of WiggleZ galaxies unfeasible, one can use existing multi-wavelength photometric datasets to better characterise this population. 

Identifying synergies between existing multi-wavelength photometric surveys and large redshift surveys like WiggleZ, is important for several reasons. Firstly, spectroscopic samples such as the WiggleZ sample serve as important calibration sets for large photometric surveys. As an example, the ongoing Dark Energy Survey (DES) will detect 300 million galaxies out to z$\sim$2 and will overlap the all-sky near infrared (NIR) VISTA Hemisphere Survey (VHS). The combination of optical and NIR data will allow photometric redshifts and SED fit parameters to be constrained for a very large number of galaxies out to z$\sim$2 \citep{Banerji:08}. Colour selected spectroscopic samples like WiggleZ may be used as \textit{training sets} but will by their nature be incomplete in certain regions of parameter space sampled by flux-limited photometric surveys such as DES and VHS. Quantifying this incompleteness in terms of the physical properties of the galaxies - i.e. understanding the types of galaxies that constitute currently available spectroscopic samples, is a useful exercise in order to better calibrate redshifts from photometric surveys. Next generation spectroscopic surveys like Big BOSS \citep{Schlegel:11}, 4MOST \citep{deJong:12} and DESpec \citep{Abdalla:12}, will in turn use the photometric surveys as the basis for target selection. Once again, understanding the physical properties of existing spectroscopically confirmed galaxies within these surveys, will help to design colour selection algorithms for new populations in the future.

In this work, we characterise the stellar masses and star formation histories of a large sample of $\sim$40,000 galaxies at 0.3$<z<$1.0, corresponding to a single field in the WiggleZ survey, and focus in particular on the subset of $\sim$12,000 of these galaxies that are also matched to the NIR UKIDSS Large Area Survey (ULAS). The advantage of long-wavelength data in the infrared is that it is less sensitive to the dust extinction in galaxies than the UV and optical. This, combined with the fact that older, more massive stars are brighter in the infrared, means that the infrared data is expected to give a more unbiased estimate of the total stellar mass in galaxies than the UV/optical \citep{Bell:03, Drory:04}. Recent studies have however claimed that the NIR actually leads to poorer stellar mass estimates for a sample of intermediate redshift galaxies from the Galaxy and Mass Assembly (GAMA) survey \citep{Taylor:11}. This worsening in the stellar mass estimates is attributed to uncertainties in the stellar population synthesis models at these wavelengths and/or uncertainties in calibrating the photometry between the optical and NIR surveys. The role of NIR photometry in constraining the stellar masses of galaxies therefore still remains open to debate. 

The stellar mass estimates themselves are useful for cosmology given that the clustering strengths of galaxy samples split by stellar mass are expected to be different - (e.g. \citealt{Coil:08}). Our aim in the current work is therefore also to test the robustness of stellar mass estimates from SED fitting codes, to various assumptions made during the fitting process. The WiggleZ galaxies are selected to be extremely blue galaxies \citep{Drinkwater:10}, and as such, represent a sample where SED fitting is likely to be the most problematic. By testing SED fitting codes on this sample, we can be reasonably confident that the codes can be applied to older, redder and more massive galaxies, which are likely to have less complex star formation histories. 

Throughout this paper we assume a flat $\Lambda$CDM cosmology with $h$=0.7. All magnitudes are on the AB system where the UKIDSS photometry has been converted to the AB system using the conversions in \citet{Hewett:06}: Y=+0.633, J=+0.937, H=+1.376 and K=+1.897. The Vega to AB conversions in the WISE bands are assumed to be W1=+2.683, W2=+3.319, W3=+5.242 and W4=+6.404 \citep{Cutri:12}.   

\section{DATA}

\label{sec:data}

We begin by describing the spectroscopic and photometric catalogues that are used in this work. 

\subsection{Wiggle-Z Dark Energy Survey}

Spectroscopic data for the UV luminous emission line galaxies is taken from the WiggleZ Dark Energy Survey \citep{Drinkwater:10}. The survey has assembled reliable redshifts for 219,682 galaxies in seven different fields. The spectroscopic targets are selected using ultraviolet data from the \textit{GALEX} satellite \citep{Martin:05} and optical data from the Sloan Digital Sky Survey (SDSS) in the north \citep{Adelman:06} and the Canada France Hawaii Telescope Red Sequence Cluster Survey in the south \citep{Yee:07}. A series of magnitude and colour cuts are applied to the data to preferentially select  star-forming galaxies with bright emission lines which are then targetted using the AAOmega spectrograph on the Anglo-Australian Telescope \citep{Sharp:06}. Full details of the spectroscopic target selection can be found in \citet{Drinkwater:10} but results in a complicated selection function over the total survey area. This selection function is computed in \citet{Blake:10} where the different sources of incompleteness are fully detailed. We note in particular that the complicated colour selection of WiggleZ targets means the sample is incomplete in some regions of redshift, stellar mass and star formation rate parameter space. 

\begin{figure}
\begin{center}
\includegraphics[width=8.5cm,height=6.0cm,angle=0] {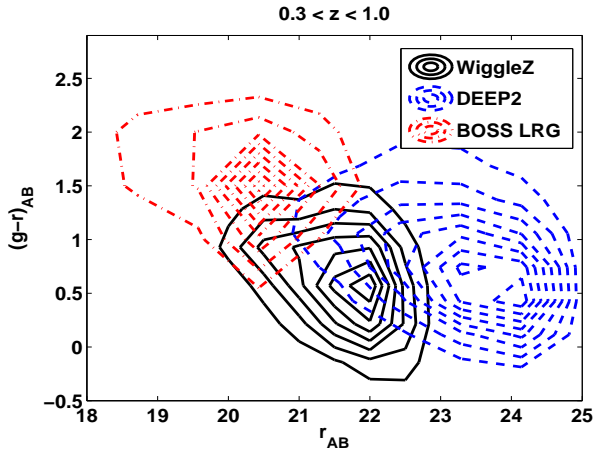}
\caption{Observed $r$-band magnitude versus $(g-r)$ colour for all WiggleZ galaxies in the 15hr field at 0.3$<z<$1.0, compared to star-forming galaxies from the DEEP2 survey and Luminous Red Galaxies from the BOSS survey, both at similar redshifts to WiggleZ. The contours represent the density of points computed using a kernel density estimator over a grid traversing the parameter space covered by the samples. The WiggleZ galaxies are brighter than the DEEP2 sample but slightly bluer than these DEEP2 galaxies which represent normal star-forming galaxies at these redshifts.}
\label{fig:select}
\end{center}
\end{figure}

In this paper, we work with the WiggleZ data in the 15hr field only with 209$<$RA$<$231 and $-$3.2$<$DEC$<$7.2 in order to ensure overlap with currently available infrared datasets from UKIDSS and \textit{WISE}. This field contains 46,144 galaxies in total down to a flux limit of NUV$<$22.8. All photometric catalogues from the WiggleZ survey contain the de-reddened galaxy magnitudes and this sample only contains sources with redshift quality between 3 and 5 which corresponds to reliable redshift estimates. We select only galaxies at 0.3$<$z$<$1.0 for this study which constitutes the bulk of the WiggleZ population. At higher redshifts, the quality of the redshift estimate becomes increasingly unreliable and many of the $z>1$ sources with reliable redshift measurements are AGN. Although these AGN may contain emission related to star formation, the SED fitting codes used in this work do not allow us to disentangle the contributions of the two. We are primarily interested in the star-formation dominated WiggleZ galaxies in this work and so we restrict our redshift range to a regime where the AGN make up an insignificant proportion of the population. Our final sample of WiggleZ galaxies therefore totals 39,701 sources at 0.3$<$z$<$1.0. The WiggleZ magnitudes used throughout this paper correspond to the de-reddened model magnitudes from SDSS and \textit{GALEX}. The WiggleZ galaxies are typically $\sim$5$\sigma$ detections in the \textit{GALEX} NUV band and $>$10$\sigma$ detections in the SDSS $r$-band.  

In order to better illustrate the types of galaxies selected using the WiggleZ colour cuts, in Figure \ref{fig:select} we show the observed colour-magnitude distribution of the WiggleZ galaxies. This is compared to $\sim$19,000 galaxies from the DEEP2 survey Data Release 4 \citep{Newman:12} over the same redshift range. These DEEP2 galaxies represent typical star-forming galaxies at these redshifts. We also compare the distribution to $\sim$180,000 Luminous Red Galaxies (LRGs) over the same redshift range selected from within the SDSS BOSS survey \citep{Maraston:12}. As expected, the WiggleZ galaxies are considerably bluer and fainter than the LRGs that make up the red sequence at these redshifts. However, despite the brighter flux limit of WiggleZ compared to DEEP2, at a fixed $r$-band magnitude, the WiggleZ selection targets sources that are also bluer than the typical \textit{blue cloud} galaxies that make up the DEEP2 sample. In other words, the WiggleZ selection is isolating the most extreme end of the blue galaxy population at any given luminosity. 


Having described the properties of the WiggleZ sample, we now move on to considering synergies between this sample and wide-field infrared (IR) imaging surveys.

\subsection{UKIDSS Large Area Survey}

The UKIDSS Large Area Survey \citep{Lawrence:07} is the current largest NIR survey and has obtained imaging over $\sim$3200 deg$^2$ of the northern sky in the $Y$,$J$,$H$ and $K$ bands. The survey is being carried out using the Wide Field Camera (WFCAM) on the 3.8m UK infrared Telescope (UKIRT). UKIDSS, which began in 2005, is the successor to the Two Micron All
Sky Survey (2MASS) and is the NIR counterpart to the Sloan
Digital Sky Survey (SDSS). We use Data Release 9 of the UKIDSS LAS (ULASDR9) in this work, which reaches nominal 5$\sigma$ depths of $Y$=20.8, $J$=20.5, $H$=20.2 and $K$=20.1. Throughout this work, we use the Petrosian magnitudes in the UKIDSS catalogues as the UKIDSS catalogues do not include model magnitudes. These Petrosian magnitudes serve as a reasonable estimate of the total NIR flux of the galaxy. Differences between the SDSS Petrosian magnitudes and SDSS model magnitudes of galaxies are of the order of 15 per cent \citep{Banerji:10}.   

We match the WiggleZ galaxies in the 15hr field to UKIDSS using a matching radius of 2$^{\prime \prime}$. The median separation between the WiggleZ and UKIDSS sources is $\sim$0.3$^{\prime \prime}$. 11,919 of the original 39,701 galaxies are detected at $>$5$\sigma$ in at least one of the UKIDSS bands corresponding to 30 per cent of the WiggleZ sample. The redshift distribution for this UKIDSS detected sub-sample is very similar to that of the entire population at 0.3$<z<$1.0 with a peak at z$\sim$0.7. The UKIDSS sub-sample is compared to those WiggleZ galaxies not matched to a NIR source in Figure \ref{fig:select2}. We find, as expected that the fraction of very blue galaxies detected in the NIR is very low and increases as we go to redder colours. The UKIDSS detected galaxies are also brighter than those that are undetected. The median $r$-band magnitude of the UKIDSS detected galaxies is 21.0 versus 21.8 for the galaxies undetected in UKIDSS. The median $(g-r)$ colour is also 0.2 magnitudes redder for the UKIDSS detected galaxies. 

\subsection{Wide-Field Infrared Survey Explorer}

The Wide-Field Infrared Survey Explorer (\textit{WISE}; \citealt{Wright:10}) has conducted an all-sky survey in four passbands - 3.4, 4.6, 12 and 22$\mu$m with 5$\sigma$ depths of $>$19.1, $>$18.8, $>$16.4 and $>$14.5. By its nature, \textit{WISE} contains many different populations of astrophysical sources including planetary debris disks, populations of cool low-mass stars and ultraluminous infrared galaxies and active galactic nuclei. Although it is very shallow at 12 and 22$\mu$m, the 3.4 and 4.6$\mu$m depths are reasonably well-matched to the UKIDSS-LAS so it is interesting to ask what fraction of very blue star-forming galaxies such as the WiggleZ sample, are detected at these wavelengths in a relatively shallow all-sky IR survey.

\begin{figure}
\begin{center}
\begin{tabular}{c}
\includegraphics[scale=0.4,angle=0] {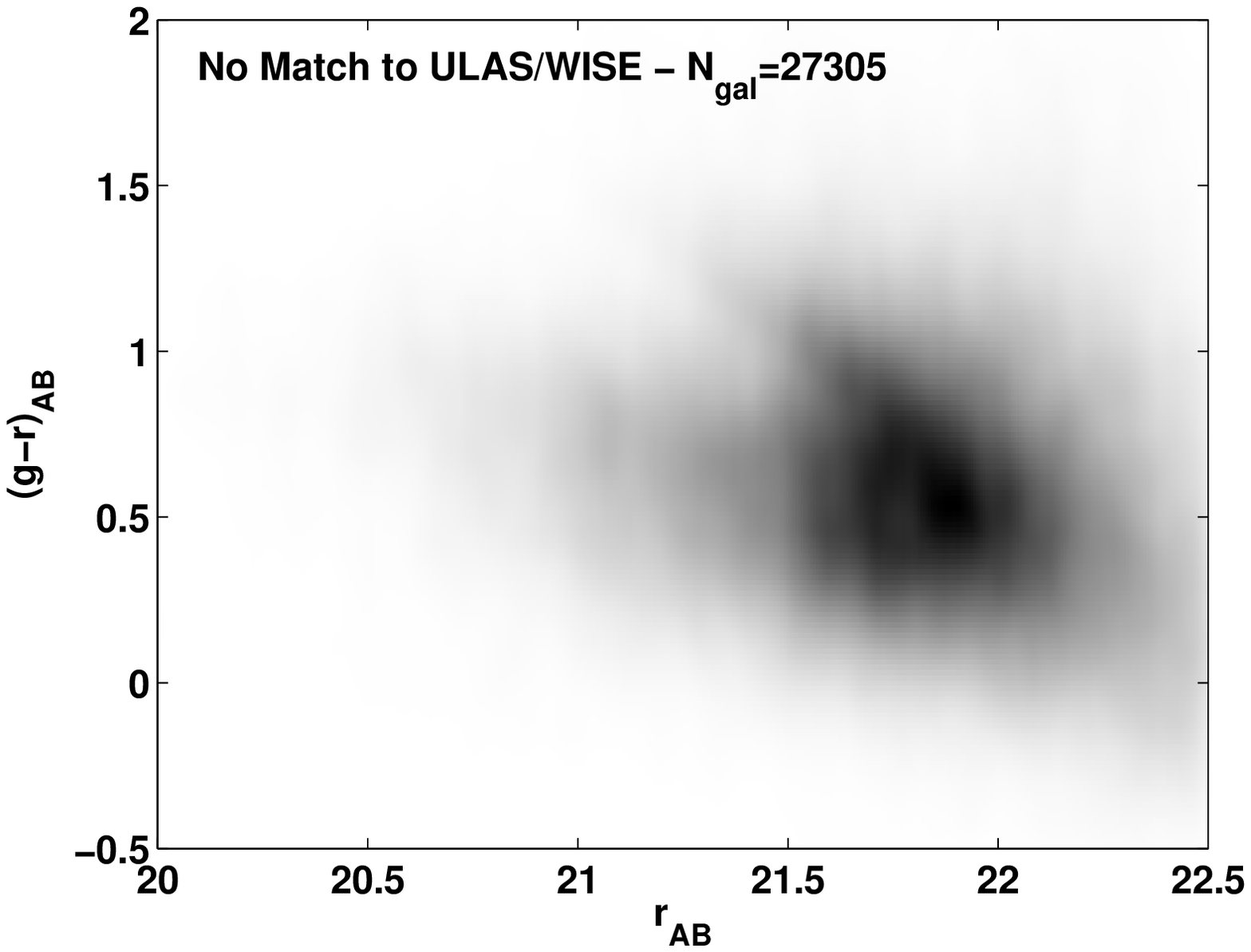} \\
\includegraphics[scale=0.4,angle=0] {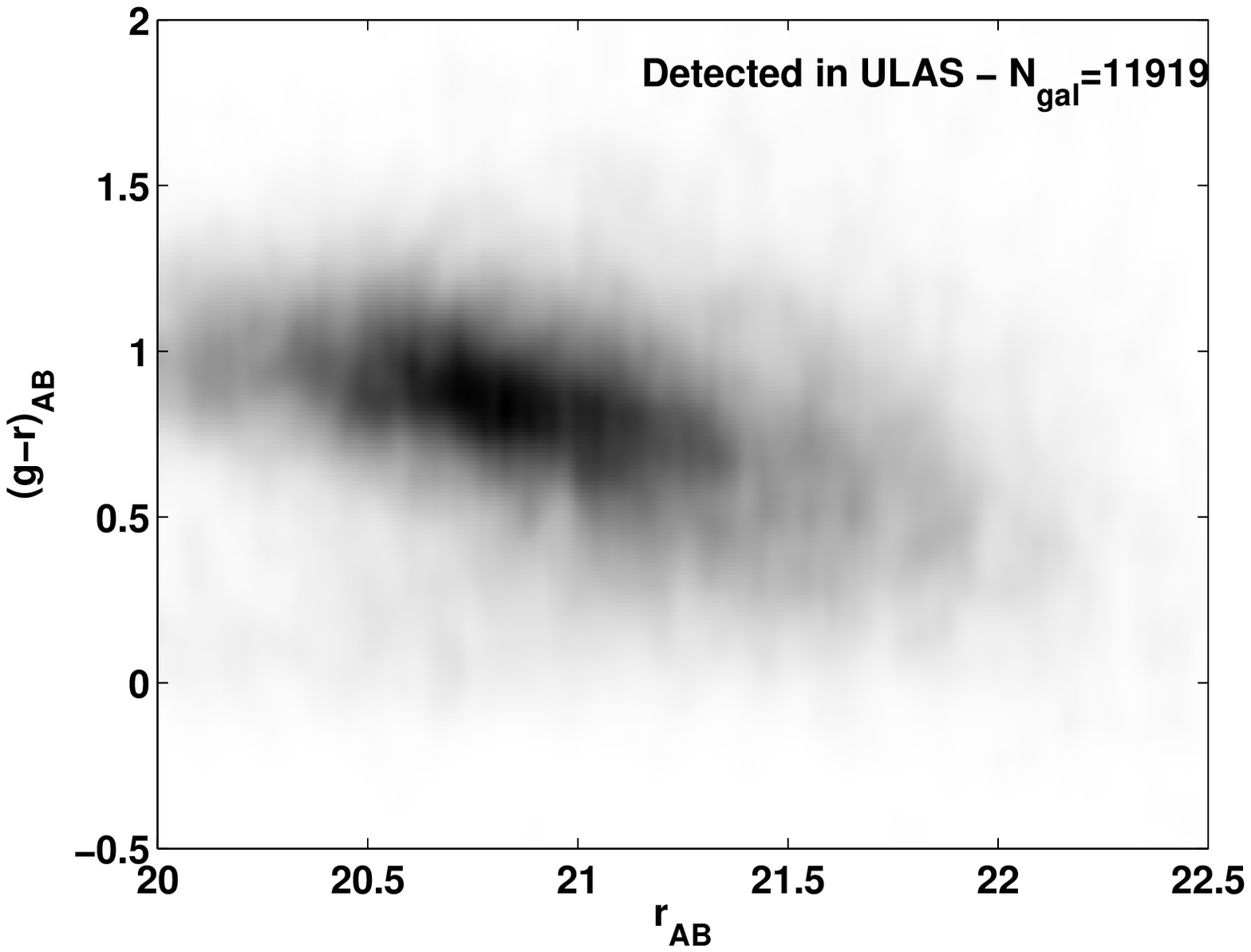} \\
\includegraphics[scale=0.4,angle=0] {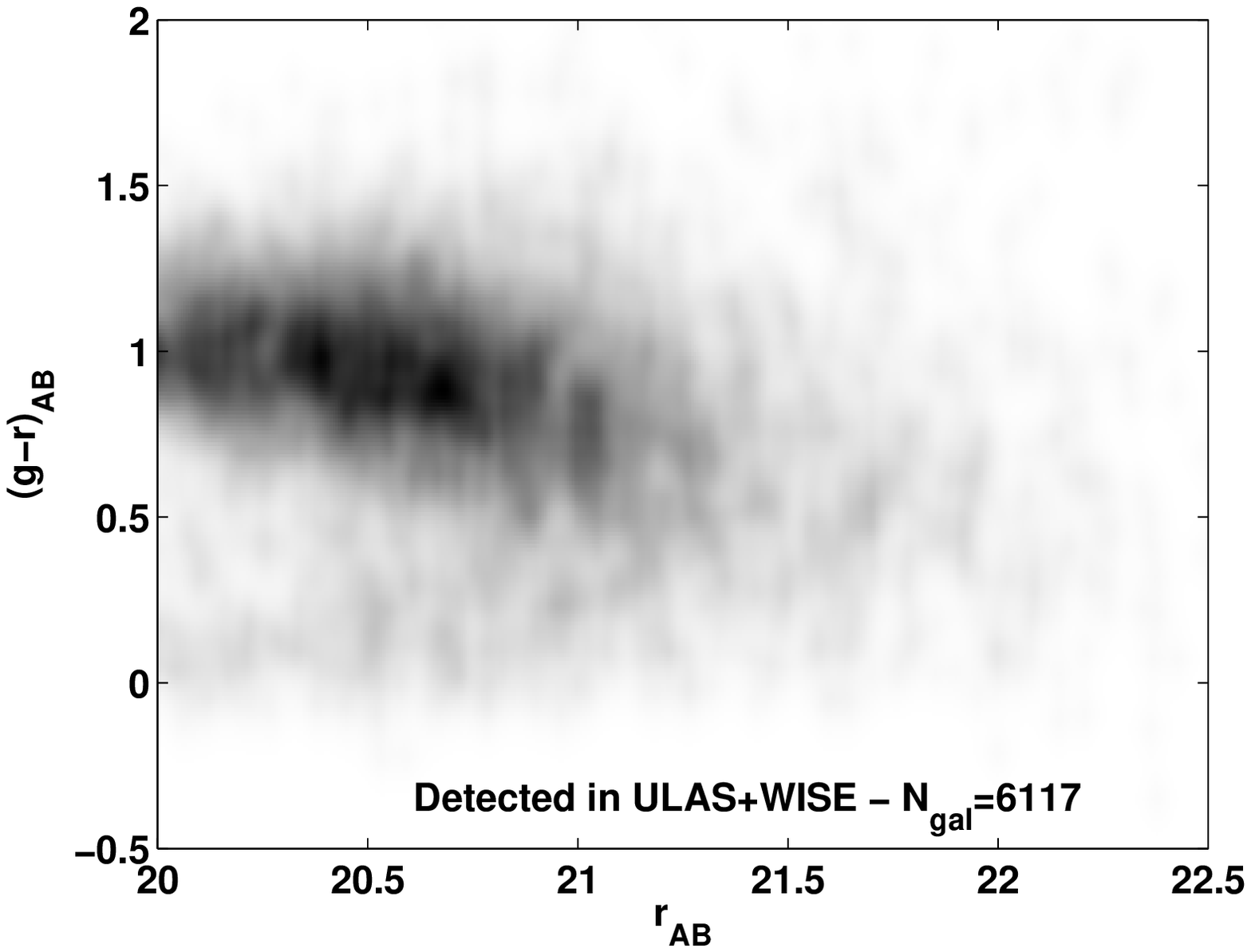}\\
\end{tabular}
\caption{Observed-frame $(g-r)$ colour magnitude diagram of infrared detected and infrared undetected WiggleZ galaxies showing that the UKIDSS-LAS and WISE detected sub-samples constitute the brighter, redder end of the WiggleZ population.  The greyscale represents the density of points. The redshift distributions for all three samples are very similar and peak at $z=0.7$.}
\label{fig:select2}
\end{center}
\end{figure}

We match our sample of 39,701 WiggleZ galaxies at 0.3$<$z$<$1.0 to the WISE All-Sky Release using a matching radius of 4$^{\prime \prime}$. The angular resolution at 3.4 and 4.6$\mu$m is $\sim$6$^{\prime \prime}$. We select only those galaxies that are detected at S/N$>$5 at 3.4$\mu$m and S/N$>$3 at 4.6$\mu$m and have \textit{WISE} colours of [3.4$_{\mu m}-$4.6$_{\mu m}]<$0.8, consistent with these galaxies not having a significant AGN component \citep{Assef:10}. At the \textit{WISE} wavelengths, the presence of an AGN can significantly affect the galaxy colours and the SED fitting codes used in this work do not allow us to fit for this AGN component. The median separation between the \textit{WISE} and WiggleZ sources is $\sim$0.5$^{\prime \prime}$ and as expected, larger than that between the UKIDSS and WiggleZ sources. Note that using a smaller matching radius of 2$^{\prime \prime}$ rather than 4$^{\prime \prime}$, decreases the number of galaxies by only $\sim$10\% and does not affect any of our results. We use the \textit{WISE} magnitudes obtained from profile fitting as detailed in \citet{Cutri:12}, which mitigates the effects of source confusion and source blending which can be significant at the typical fluxes of our sample. At the redshifts of the WiggleZ sample, we note that most of the galaxies appear unresolved in the \textit{WISE} images. More than 90\% of the \textit{WISE} matches are also in UKIDSS. Visual inspection of those that are not shows them to mostly be blended sources where two nearby sources that are resolved separately in the UKIDSS images, are blended together in \textit{WISE}. The remaining \textit{WISE} only identifications correspond to small patches of sky with no UKIDSS coverage. We therefore only restrict our \textit{WISE} sample to those sources that are also present in UKIDSS. The \textit{WISE}+UKIDSS sub-sample contains 6117 galaxies at 0.3$<z<1.0$. Their colour-magnitude distribution can also be seen in Figure \ref{fig:select2}. The \textit{WISE} sample is brighter and slightly redder than the UKIDSS sample and corresponds to the brightest and reddest end of the WiggleZ population. The median $r$-band magnitude for the \textit{WISE} detected galaxies is 20.7 and these galaxies are 0.05 magnitudes redder on average than the UKIDSS detected galaxies. Once again we find very little dependence of the matched fraction on the source redshift with the \textit{WISE} sample also peaking at z$\sim$0.7.


We also note that, despite the very shallow flux limit in the \textit{WISE} 12 and 22$\mu$m band, there are 78 WiggleZ galaxies that are detected at $>$5$\sigma$ at these longer wavelengths. Thirty-seven of these also have [3.4$_{\mu m}-$4.6$_{\mu m}]<$0.8, consistent with these galaxies not having a significant AGN component. A further sub-sample of 24 of these satisfy the more conservative colour cut of [3.4$_{\mu m}-$4.6$_{\mu m}]<$0.7 \citep{Stern:12} and can reasonably be assumed to be star-formation dominated. We characterise the SEDs of these UV-luminous mid-IR bright sources in Section \ref{sec:ir}.  The remaining WiggleZ galaxies with mid infrared emission can reasonably be thought to be AGN dominated and are not considered further in this work. 

\section{Spectral Energy Distribution (SED) Fitting}

\label{sec:sed}

A key aim of this work is to derive SED fit parameters and in particular stellar masses, for the various sub-samples of UV-luminous galaxies from WiggleZ and to test the robustness of these parameters to assumptions made during the fitting process. In particular, we would like to test how these various assumptions affect the stellar mass estimates as a function of the UV-luminosity. We therefore begin by describing the various SED fitting codes that are used in this work and highlight similarities and differences between them.  

\subsection{Fitting and Analysis of Spectral Templates (FAST)}

The FAST code is fully described in \citet{Kriek:09} and is designed to fit a range of different stellar population models to fluxes of galaxies in the optical and NIR bands. It has been succesfully applied to infer the global properties of NIR selected samples of galaxies out to z$\sim$2 in the NEWFIRM medium band survey \citep{Kriek:10}. Currently FAST allows fitting of both \citet{BC:03} - BC03 hereafter - and \citet{Maraston:05} models at a range of metallicities and using different initial mass functions (IMFs) and single component star formation histories (SFH) such as exponentially decaying SFHs, delayed exponentially decaying SFHs and truncated SFHs. The BC03 models can be used in conjunction with the \citet{Salpeter:55} or \citet{Chabrier:03} IMF, while the Maraston model libraries are constructed using both the \citet{Salpeter:55} and \citet{Kroupa:01} IMFs. Interstellar dust extinction is accounted for using the \citet{Calzetti:00} extinction law and is allowed to vary over a range of A$_V$. 

The advantage of FAST is that once a grid of model fluxes has been computed, the SED fitting is relatively quick as the code uses a Monte Carlo sampling of the model parameter space to determine the best-fit parameters for each observed galaxy. This Monte Carlo sampling also allows FAST to calibrate the confidence intervals for each SED fit parameter that is estimated. Errors on the SED fit parameters are calibrated using 100 Monte Carlo simulations. Therefore, the code provides realistic errors on the SED fit parameters that take into account both the uncertainties in the broadband fluxes used for the fitting as well as the uncertainties in the SED models via the use of a template error function \citep{Brammer:08}. These errors along with the reduced $\chi^2$ metric output by the code, can be used to assess the quality of the fits for different choices of input parameters.  

FAST is readily applicable to the large sample sizes currently being assembled in cosmological volume surveys due to the speed with which it computes the best-fit SEDs. It does not however currently allow the inclusion of episodic bursts of star formation and the galaxy templates included do not contain emission lines which may be important for our sample of UV luminous galaxies. FAST also does not include inverse-$\tau$ or exponentially increasing star-formation histories which may be more representative of actively star-forming galaxies \citep{Maraston:10}. Finally, we note that the SED models used by FAST do not account for dust emission in the rest-frame infrared and assume that the SEDs of the galaxies over the rest-frame wavelength range probed by our data, are dominated by starlight. Recent studies have however noted the presence of excess emission at near infrared wavelengths that cannot be accounted for simply by the stellar continuum. This excess emission is best modelled as an additional greybody component with a temperature of between 750K and 1200K (e.g. \citealt{daCunha:08}). Some studies have found that the emission correlates with star formation and have therefore attributed it to emission from circumstellar disks around massive stars \citep{Mentuch:09, Mentuch:10}. However, as noted in these studies, this excess dust emission above the stellar continuum only starts to contribute at rest-frame wavelengths above $\sim$2$\mu$m.

 \subsection{MAGPHYS}
 
 \label{sec:magphys}

The MAGPHYS code \citep{daCunha:08, daCunha:11} is designed to consistently treat the combined UV, optical and infrared emission from galaxies. In this code, any attenuation of starlight by dust at bluer wavelengths appears as reprocessed thermal emission in the infrared where the dust is treated in a physically consistent way between the different wavelengths. The code uses the updated \citet{Bruzual:07} model SEDs which include a new prescription for the TP-AGB stars to better reproduce the NIR colours of intermediate age stellar populations. An advantage over FAST is that the code also allows for random bursts of star formation to be added to the simple exponentially decaying SFH with bursts occurring with equal probability at all times since the formation redshift. The burst probability is set so that up to 50\% of the galaxies in the model library have experienced a burst within the last 2 Gyr. The fraction of stellar mass formed in the burst ranges between 0.03 and 4$\times$ that formed through continuous star formation with a characteristic timescale $\tau$. A consequence of the facility to incorporate more complex star formation histories as well as the facility to compute the IR emission in the galaxies in a self consistent way, is that the code is also slower. MAGPHYS compares the observed photometry of galaxies to a total of $\sim$661 million models and the priors imposed on the parameters are deliberately not overly restrictive so as to ensure that the entire multi-dimensional observational space is reasonably well sampled. However this very large parameter space also means that the code is not as easily applicable to very large samples of galaxies such as the one in this paper. 

As the main advantage of this code over the others, is in the consistent modelling of the UV to infrared emission in galaxies, we only use MAGPHYS to characterise the properties of the small subset of WiggleZ galaxies that are bright at 12 and 22$\mu$m as well as to carry out independent checks on some of our results obtained using the other codes, using smaller random sub-samples of the WiggleZ galaxies. 

\subsection{KG04 Code}

We also use a non-public code developed by one of us (KG) in the Perl Data Language\footnote{http://pdl.perl.org}, which has been used in several papers (e.g. \citealt{Glazebrook:04, Baldry:08}). We will refer to this code as the `KG04 code'. This code uses the PEGASE.2 SPS models \citep{Fioc:97, Fioc:99} and incorporates two-component star-formation histories with a
primary (long term) SFH plus a burst component. The primary component is represented by an exponentially decaying burst (with values $\tau=0.1, 0.2, 0.5, 1,2,4,8,500$ Gyr)\footnote{$\tau=500$Gyr is intended to approximate constant SFR models.} and the burst component is represented as a rapid exponential SFH of $\tau=100$ Myr. The burst is allowed to occur over the full range of a galaxy SFH and is allowed to contribute a range of final stellar mass ratios from zero up to twice the long-SFH mass (evaluated at 13 Gyr). One important difference from the MAGPHYS code (see Section \ref{sec:magphys}) is that the burst fraction and times are not random but distributed across a full grid range.
Dust is allowed to vary over a range of A$_V$ with a Calzetti law, metallicity is held fixed with time in the PEGASE.2 models but is allowed to vary in the fitting. Nebular emission lines can optionally be included although we note that the line ratios are held fixed in the models \citep{Fioc:99}. 

In a similar fashion to FAST the computation is sped up by pre-computing grids of model photometry values at each redshift of interest, and mass values and errors of a particular galaxy are then determined by a fast minumum-$\chi^2$ lookup of the photometry with a Monte-Carlo realisations of the errors. The final mass and error values are the mean and standard deviation of five Monte-Carlo realisations.
The grids in the various parameters are stepped in a pseudo-logarithmic fashion because of (i) the optimisation of the sampling of each parameter with respect to physical variations and (ii) the constrained values of certain parameters such as the fixed metallicity values of the PEGASE.2 models. A total of 2,427,480 possible model SEDs are considered for each galaxy in the fitting (with some being automatically excluded by the
age of the Universe at a given redshift constraint). Normally the code has been used with the \citet{Baldry:03} IMF (BG03 hereafter), however it can be adapted to any IMF as is done in this paper.

Stellar masses are the masses locked up in luminous stars at the best fit time but exclude mass contributions from non-luminous stellar remnants: white dwarfs, neutron stars and black holes. 

\section{RESULT OF SED FITTING}

\label{sec:results}

In this section, we test the robustness of stellar mass estimates for the WiggleZ galaxies to various assumptions made during the SED fitting procedure and consider in particular the dependence of the mass estimates on the rest-frame $FUV$ luminosity.  

\subsection{The Effect of Photometric Bands}

We begin by assessing the role of infrared photometry in constraining the stellar masses of the WiggleZ galaxies. There are several reasons why infrared data is expected to help in constraining the total stellar mass of a galaxy. Firstly, the rest-frame NIR is less sensitive to dust extinction than the UV and optical and therefore provides a more unbiased view of stars in galaxies. Secondly, the most massive evolved stars (i.e. those that persist on Gyr timescales) in galaxies are preferentially redder and emit strongly in NIR wavelengths so the infrared light provides a better representation of the high-mass end of the stellar mass function which will contribute significantly to the total stellar mass. The UV/optical wavebands are dominated by light from young luminous short-lived stars which contribute significantly less to the total stellar mass. Finally, and particularly relevant for our sample, in spectroscopic surveys covering a wide redshift range, the availability of additional photometry at longer wavelengths allows us to sample the same rest-frame portion of the galaxy SED at high redshifts as sampled by the UV and optical filters at low redshifts. For these reasons, rest-frame NIR mass-to-light ratios have often been used to derive robust stellar masses for galaxies \citep{Bell:03, Drory:04}. Although it is true that stellar population synthesis (SPS) models suffer from larger uncertainties in modelling the rest-frame NIR SEDs of galaxies, it is important to quantify the effect these uncertainties have on the stellar masses and how these trade off with the inclusion of additional data points in the fitting procedure. 

We use the FAST code to derive stellar masses for a) the 27,305 WiggleZ galaxies unmatched to wide-field IR surveys ($FUV,NUV,ugriz$ photometry), b) the 11,919 WiggleZ galaxies at 0.3$<$z$<$1.0 that are also detected at $>$5$\sigma$ in at least one of the UKIDSS bands ($FUV,NUV,ugrizYJHK$ photometry) and c) the 6117 galaxies in the same redshift range that are also detected in \textit{WISE} ($FUV,NUV,ugrizYJHK,W1,W2$ photometry). About a quarter of the 11,919 UKIDSS matched galaxies are detected in all four UKIDSS bands - $Y,J,H,K$. In some cases, data is missing from one or more of the UKIDSS bands due to the fact that there are regions of sky that haven't been imaged in all four bands. However, $\sim$20\% of the galaxies are only detected in the $Y$-band on account of being very blue and fall below the S/N threshold used to construct the UKIDSS catalogues in the redder UKIDSS bands. For this subset, it is also interesting to assess the effect of using the low S/N fluxes in the red bands in the SED fits versus ignoring the galaxy photometry in these bands. 

In all cases we assume BC03 SPS models, an exponentially decaying star formation history (SFH) with $\tau$ ranging from 0.01 to 30Gyr - the latter representing constant star formation - a grid of metallicities with 0.001$<$Z$<$0.05 and a \citet{Chabrier:03} IMF. Dust extinction is allowed to vary over 0$<$A$_V<$2. We will test how some of these parameter choices can affect the stellar mass estimates, later in this section. 

\subsubsection{Differences in IR undetected, UKIDSS detected and UKIDSS+\textit{WISE} detected sub-samples}

\label{sec:ir1}

\begin{figure}
\begin{center}
\begin{tabular}{ccc}
\includegraphics[scale=0.35,angle=0] {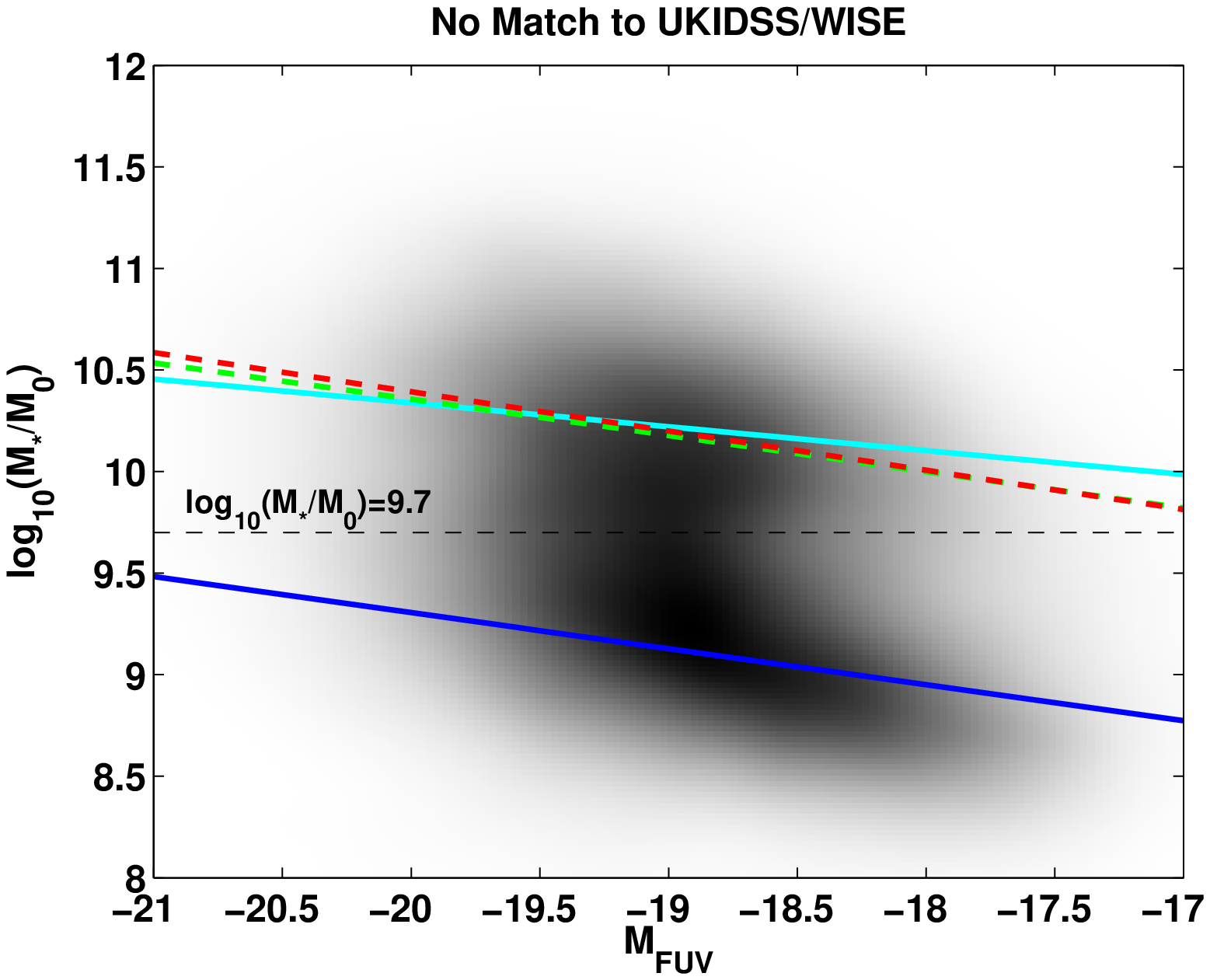} \\
\includegraphics[scale=0.35,angle=0] {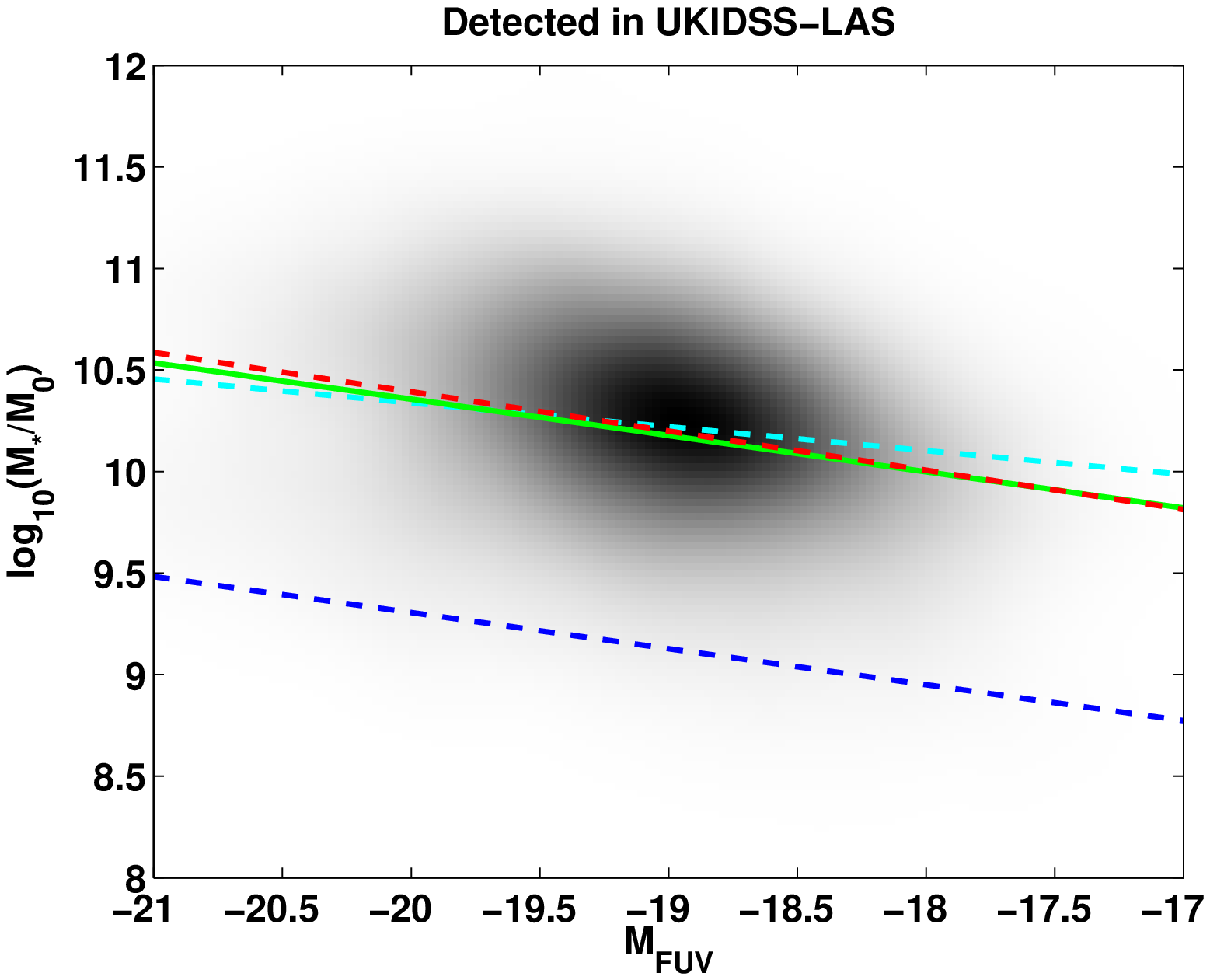} \\
\includegraphics[scale=0.35,angle=0] {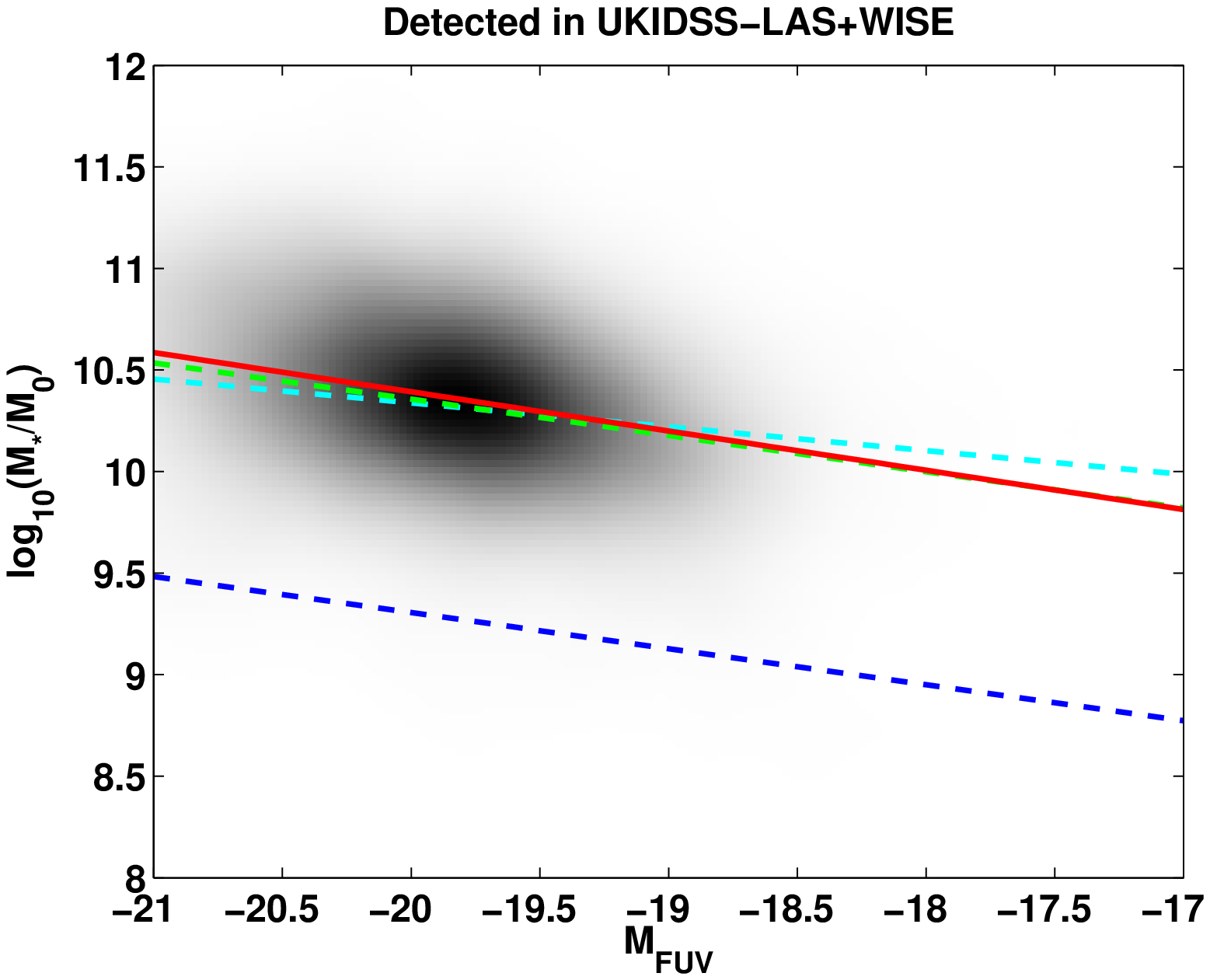}\\
\end{tabular}
\caption{Stellar mass as a function of $FUV$ absolute magnitude for the WiggleZ galaxies not detected in wide-field IR surveys (top), those detected in UKIDSS out to 2.2$\mu$m (middle), and those detected in both UKIDSS and \textit{WISE} out to 4.6$\mu$m (bottom). The greyscale represents the density of points in each panel. The galaxies detected in the IR have a much tighter stellar mass distribution compared to the bluer galaxies missed in wide-field IR surveys. They are also clearly more massive. In each panel, we show the best-fit straight line through these points as the solid line to guide the eye and the best-fit lines for the other panels are shown as the dashed lines for comparison. For the top panel of galaxies undetected in the IR, we have fit two separate best-fit lines to the two clouds of galaxies at log$_{10}$(M$_*$/M$_\odot$)$<9.7$ (blue line) and log$_{10}$(M$_*$/M$_\odot$)$>9.7$ (cyan line).}
\label{fig:nir_mass1}
\end{center}
\end{figure}

In Figure \ref{fig:nir_mass1}, we plot the stellar mass distributions for the IR-missed, UKIDSS and UKIDSS+\textit{WISE} samples as a function of the $FUV$ absolute magnitude. The $FUV$ absolute magnitude used throughout this paper is derived by k-correcting the observed $NUV$ magnitude for each galaxy using a Lyman Break Galaxy (LBG) template reddened by A$_V$=0.14 mags. This LBG template produces a very good match to the average observed $(NUV-r)$ colours of the WiggleZ galaxies \citep{Blake:09}. Some of the WiggleZ galaxies, particularly those detected in the IR, are redder than this template, but our best-fit SEDs only begin to diverge from the LBG template at rest-frame wavelengths of $\lambda \gtrsim$2000\AA\@. The $FUV$ luminosity is found to be relatively insensitive to the choice of template and we find that differences in the $FUV$ luminosity computed using this LBG template versus the individual best-fit SEDs, are at most $\sim$10\%. As we will be deriving stellar masses and SED fit parameters using a range of different inputs, when considering how these fit parameters change as a function of rest-frame galaxy luminosity or colour, it is important to ensure that the rest-frame quantity used does not depend critically on the form of the best-fit SED. For this reason, the $FUV$ luminosity is used throughout this paper. Also, for ease of computation, the single LBG template is used to derive the $FUV$ k-corrections and we have checked that using the individual SEDs instead does not affect any of our conclusions.

We find that the IR detected WiggleZ galaxies have a fairly tight distribution in stellar mass while the less luminous galaxies that are unmatched to wide-field IR surveys, span a much larger range in stellar mass and have in general lower stellar masses and larger errors on the individual stellar mass estimates. The median stellar masses are  log$_{10}$(M$_*$/M$_\odot$)=9.6$\pm$0.7, 10.2$\pm$0.5 and 10.4$\pm$0.4 for the IR missed, UKIDSS matched and UKIDSS+\textit{WISE} matched sub-samples respectively. It is important to clarify the meaning of the various errors on the stellar masses quoted throughout this paper. The quoted errors on the median masses, refer to the standard deviation for the entire sample. The median 1$\sigma$ errors on the individual mass estimates, derived using Monte Carlo simulations, is typically slightly smaller - $\sim$0.3--0.4 dex. The error on deriving this median on the other hand, assuming Gaussian statistics, is extremely small given our large sample size ($<$0.01 dex). From hereon, the errors on the median reflect the standard deviation of the sample, but we will also refer to the 1$\sigma$ errors computed from the Monte Carlo simulations as a means of assessing the accuracy of the stellar mass estimates. 

We also note that the IR undetected galaxies seem to have a bimodal distribution in the stellar mass - M$_{FUV}$ plane with a cloud of galaxies with similar stellar masses and $FUV$ luminosities to the IR-detected galaxies, and a separate cloud of fainter, lower stellar mass galaxies. There is a clear separation between these two populations at a stellar mass of log$_{10}$(M$_*$/M$_\odot$)$\sim9.7$. The lower stellar mass galaxies are predominantly at lower redshifts whereas at higher redshifts, the WiggleZ colour cuts seem to select more massive galaxies. We note however that the IR detected population, which is predominantly more massive than log$_{10}$(M$_*$/M$_\odot$)$\sim9.7$, contains galaxies over the entire redshift range of the WiggleZ survey. Using the IR detected galaxies, we check that excluding the IR data from the SED fitting, does not reproduce this observed bimodality. We find therefore that at the high redshift end of the WiggleZ sample, there is a significant population of massive galaxies that are undetected in wide-field NIR surveys, whereas at the low redshift end of the sample, those galaxies that are undetected in the NIR, are predominantly fainter and have lower stellar masses. We note further that this bimodality is also somewhat mirrored in the observed colour-magnitude diagram in Figure \ref{fig:select2} where we see that some of the WiggleZ galaxies that are undetected in the NIR surveys, are just as red in terms of their optical colours as the NIR detected galaxies.

Reliable age constraints are difficult to obtain from these SED fits. For single component star formation histories used in the FAST code, the age refers to the age of the galaxy since the onset of star formation, and ranges from $\sim$250 Myr for the IR-undetected population, with typical 1$\sigma$ errors of 2--3 Gyr on these age estimates, versus $\sim$400--1000 Myr $\pm$1.5 Gyr for the IR detected galaxies. The age is very degenerate with the dust extinction with both parameters having a similar effect on the observed colours of galaxies. The UKIDSS and \textit{WISE} detected galaxies are found to have higher A$_V$ by $\sim$ 0.4 magnitudes, compared to the galaxies not detected in the infrared. However, the typical 1$\sigma$ errors on these A$_V$ estimates are $\sim$0.6 magnitudes so the age and dust extinctions are not particularly well constrained.    

\subsubsection{How does the IR data change the stellar masses?}

\begin{figure}
\begin{center}
\begin{tabular}{c}
\includegraphics[scale=0.4,angle=0] {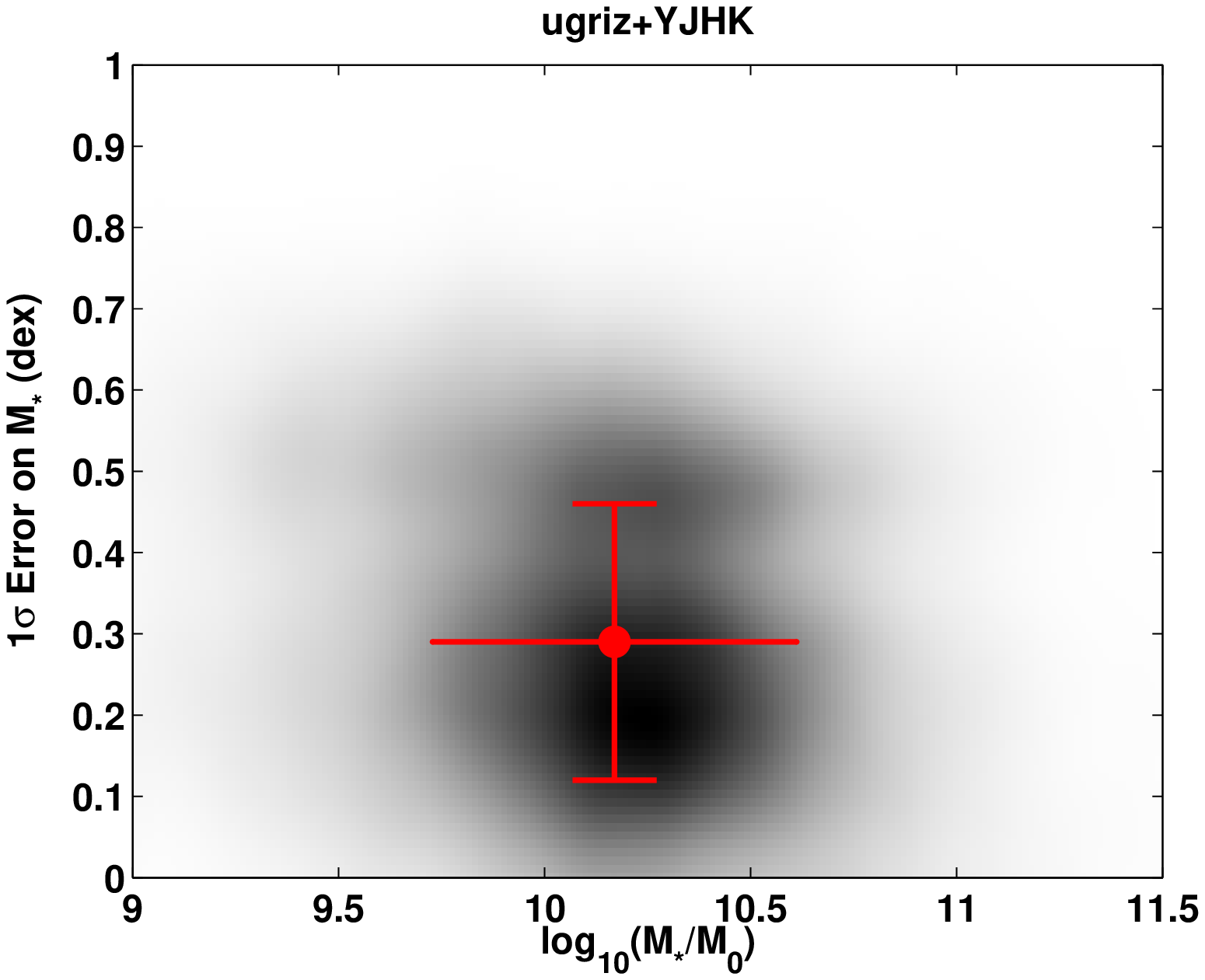} \\
\includegraphics[scale=0.4,angle=0] {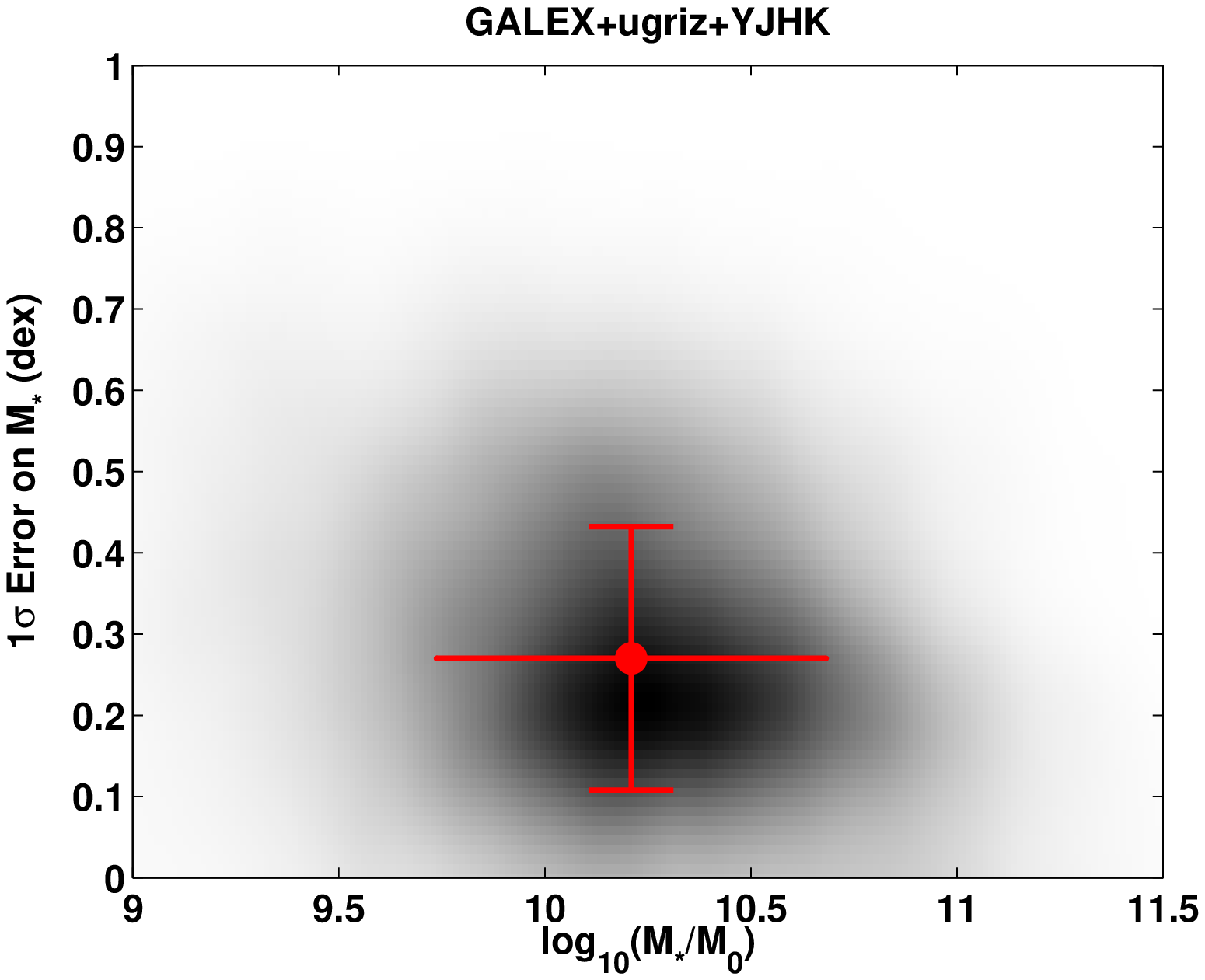} \\
\includegraphics[scale=0.4,angle=0] {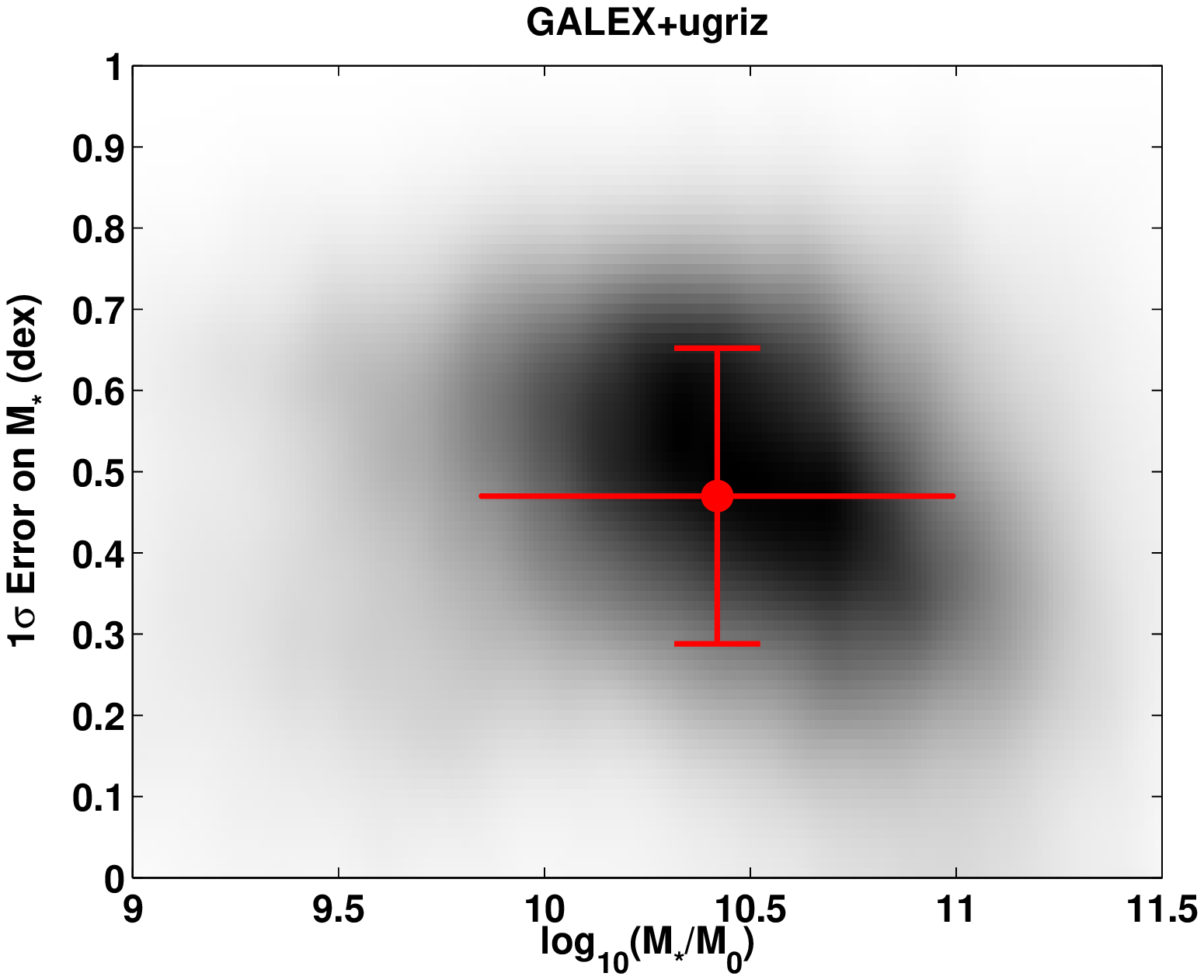}\\
\end{tabular}
\caption{The distribution of 1$\sigma$ error on the stellar mass estimates as a function of stellar mass, computed using FAST, for the sample of 11,919 WiggleZ galaxies matched to UKIDSS when using different photometric bands in the SED fitting. The greyscale represents the density of points in each panel while the points with error bars represent the median and standard deviations of the samples. The 1$\sigma$ errors are systematically larger when the NIR data is excluded from the fits (bottom panel) confirming that the inclusion of this data where available, provides better constraints on the stellar masses (middle panel). The stellar masses too are over-estimated when the NIR data is excluded from the fits. The \textit{GALEX} UV-bands however make little difference to the inferred median stellar mass and errors (top and middle panels) although there is evidence for a cloud of galaxies where the stellar masses are more poorly constrained without the UV data. This cloud of galaxies is seen to migrate from the top left of the top panel to the bottom right of the middle panel in the figure.}
\label{fig:nir}
\end{center}
\end{figure}

Are these discrepancies between the stellar masses of the IR undetected and detected population, due to differences in the physical properties of these galaxies? Or do they simply result from including different photometric bands in the fits for the two samples? In order to assess this, we compute masses for the 11,919 galaxies matched to UKIDSS, but excluding the NIR data from the SED fitting - i.e. we want to look at the effect of using only the UV/optical photometry in the SED fits for the exact same galaxies. The 1$\sigma$ errors on the stellar mass estimates, derived using FAST, are plotted as a function of the stellar mass in the bottom two panels of Figure \ref{fig:nir} both when including and excluding the NIR data from the SED fits for the subset of 11,919  galaxies. 

Figure \ref{fig:nir} clearly demonstrates that the stellar masses are better constrained with considerably smaller 1$\sigma$ errors when we include the NIR data in the SED fits in the FAST code. For the 11,919 UKIDSS detected galaxies, the median stellar mass is log$_{10}$(M$_*$/M$_\odot$)=10.2$\pm$0.5 when the UKIDSS data is included in the fitting versus log$_{10}$(M$_*$/M$_\odot$)=10.4$\pm$0.6 when the UKIDSS data is excluded. Once again we note that these errors refer to the total standard deviation for the sample and that the error on the medians are considerably smaller, making these differences in the two mass estimates highly statistically significant. 
As the removal of NIR photometry from the SED fits only serves to make the median mass of the NIR detected population even larger, we conclude that the difference in mass seen between the NIR detected and undetected galaxies in Section \ref{sec:ir1}, is real and not an artefact of including different bands in the SED fitting. The NIR detected WiggleZ galaxies are therefore on average more massive than those not detected in the infrared. 

We now turn our attention to considering whether the addition of the NIR has indeed led to better constraints on our stellar masses as suggested by Figure \ref{fig:nir}. Recently, \citet{Taylor:11} have argued that including near infrared data from the UKIDSS-LAS when SED fitting to galaxies in the GAMA survey, results in stellar mass estimates that are highly discrepant between the optical and optical+NIR fit cases. They find that the optical+NIR derived masses are inconsistent with the optical only masses at the $>$3$\sigma$ level for $\sim$25\% of the GAMA galaxies. They conclude that the NIR can therefore not be used to provide reliable stellar mass constraints for their dataset. We conduct a similar test in the case of our WiggleZ galaxies using the FAST outputs for our stellar masses. Taking into account the formal 1$\sigma$ errors on the individual mass estimates derived using the Monte Carlo simulations, we look at the fraction of galaxies where the masses from the optical and optical+NIR runs, are consistent within these 1$\sigma$ errors.  We find that this is true in $>$70\% of the galaxies indicating that although there is a systematic shift in the median mass towards lower masses when including the NIR data, the masses thus obtained are still consistent with the larger errorbars derived using the UV and optical data only. The fraction of galaxies with $>$3$\sigma$ discrepant mass estimates between the optical and optical+NIR fits, is only $\sim$4\% in the case of the FAST outputs compared to the $\sim$25\% reported in \citet{Taylor:11} for the GAMA galaxies. We therefore find that the NIR data does help better constrain the SED models for the majority of the WiggleZ galaxies.

Can the differences between our conclusions and those of \citet{Taylor:11}, be explained by FAST's use of a template error function to model uncertainties in stellar population synthesis models over certain wavelength ranges? This template error function is lowest in the rest-frame optical and increases both at rest-UV wavelengths due to the effects of varying dust extinction, and in the NIR due to uncertainties in the stellar isochrones \citep{Maraston:05}. We recompute stellar masses for the UKIDSS detected galaxies with and without the near infrared data, but this time \textit{turning off} the template error function within FAST. The template error function increases the photometric uncertainties at $\lambda \gtrsim 10000$\AA\@ \citep{Brammer:08} and therefore the resulting errors on the masses. Once again, we find that the stellar masses are more tightly constrained with the inclusion of the NIR data. The fraction of galaxies with $>$3$\sigma$ discrepant mass estimates between the optical and optical+NIR fits now goes up to $\sim$6\%, still considerably smaller than the 25\% reported by \citet{Taylor:11}.  Our improved stellar mass constraints with the NIR data, are primarily driven by the fact that the 1$\sigma$ upper bound on the stellar mass is brought down by 0.4--0.5 dex with the inclusion of the NIR photometry. The lower bound on the stellar mass computed using the Monte Carlo simulations, however remains unchanged. The NIR photometry therefore helps to rule out very large stellar masses for the majority of WiggleZ galaxies and thereby tightens the stellar mass estimates. We also check that using fewer Monte Carlo simulations to calibrate the error estimates, leads to larger discrepancies ($\sim$6\% of galaxies with inconsistent masses at the $>$3$\sigma$ level) between the optical and optical+NIR mass estimates. This is to be expected given that as we decrease the number of Monte Carlo simulations, the errors also become less reliable.

We check our results using the KG04 code used in conjunction with the PEGASE.2 SPS models. The KG04 code makes different assumptions regarding the priors on the input parameters for the SED fits and a key motivation of the current work is to test the robustness of stellar mass estimates to these differences in SED fitting codes. In contrast to the results obtained using FAST, we find little difference in both the median stellar mass estimate and the 1$\sigma$ error on it with and without the addition of the NIR photometry. We use the KG04 code with both single component star formation histories and the addition of secondary bursts on top of the smooth underlying SFH. In both cases, the median stellar mass remains unchanged on addition of the NIR data, as does the typical 1$\sigma$ error. More than 80\% of galaxies now have stellar masses that are consistent between the optical and optical+NIR SED fits within the 1$\sigma$ errors. Less than 2\% have stellar masses that are discrepant at the $>$3$\sigma$ level.  

Our study has therefore illustrated two key points. The effect that the NIR data has in constraining the stellar masses of galaxies, depends on the assumptions made in the model being fit. Hence, we get slightly different results with the FAST and the KG04 code although it is important to stress that with the NIR photometry included, the median masses produced by the two codes are very similar. Secondly, it is important to have well calibrated errors on the stellar masses that take into account both the error on the photometry as well as the error in the templates in the NIR. If we do not include the template error function within FAST the discrepancy between the optical and optical+NIR computed masses increases. Similarly, we found that if we use a smaller number of Monte Carlo simulations to calibrate the errors, the discrepancy once again increases. We conclude that the NIR data can for some models lead to tighter constraints on the stellar masses, and find no evidence for widely discrepant stellar masses computed using the optical only and optical+NIR data with either the FAST or KG04 code. 

\subsubsection{Effect of adding in low signal-to-noise fluxes}

There are a significant number of blue galaxies in WiggleZ that are detected in the UKIDSS $Y$-band but are faint and at low S/N in the redder UKIDSS bands. In these cases, we have so far ignored the $JHK$ photometry from UKIDSS in the SED fitting. We now assess whether including these low S/N flux measurements results in any improvement in the SED fits and any difference to the inferred fit parameters. In order to do so, we select a subset of $\sim$1000 galaxies that are detected in the $Y$-band but undetected in $H$ and $K$. We perform forced photometry on these galaxies in the $H$ and $K$-bands using fixed apertures centred on the WiggleZ positions. The fluxes and errors are measured in several different apertures and using the high S/N detections, we find that a 4$^{\prime \prime}$ aperture provides a reasonable match to the UKIDSS catalogue Petrosian magnitudes of the galaxies. At low signal-to-noise however, a larger aperture often results in more noisy photometry so we choose the aperture flux estimate with the highest signal-to-noise ratio to represent the total flux from the galaxy. These aperture sizes range from 2 to 5$^{\prime \prime}$.  The median S/N is $\sim$3 in both the $H$ and $K$-bands. Stellar masses are then calculated using FAST but now including the results from the forced photometry in the $H$ and $K$ bands. In $\sim$60\% of the galaxies, we find an improvement in the 1$\sigma$ stellar mass errors by a factor of $\sim$2, when including the forced photometry fluxes. However, the inferred stellar masses themselves change only by $\sim$0.01 dex. In $\sim$5\% of the galaxies, the inclusion of the forced photometry fluxes considerably worsens the SED fits. These galaxies are, as expected, typically $<1\sigma$ detections in the $H$ and $K$-band images, and the inclusion of these very noisy fluxes worsens the quality of the fits. We conclude that using low signal-to-noise aperture fluxes versus ignoring the photometry in these bands, does not materially change the stellar mass estimates for the majority of the galaxies although it can help to reduce the errors on the individual stellar mass estimates. 

\subsubsection{How does the UV data change the stellar masses?}

We also consider the effect of removing the \textit{GALEX} UV photometry from the SED fits for the 11,919 galaxies matched to UKIDSS. The corresponding stellar masses and errors computed using FAST and the BC03 models, are plotted in the top panel of Figure \ref{fig:nir}. We find that the median stellar mass goes down by only $\sim$0.02 dex versus when the \textit{GALEX} photometry is included (middle panel of Figure \ref{fig:nir}), and the median error on this stellar mass remains unchanged. More than 80\% of the galaxies have stellar masses that are consistent within the 1$\sigma$ errors when considering the outputs with and without the UV photometry.  

However, we find in Figure \ref{fig:nir} that there is a cloud of galaxies where the removal of the \textit{GALEX} photometry actually leads to larger errors on the stellar mass estimates. We investigate this cloud of galaxies in more detail. Without the UV photometry, the best-fit SEDs suggest that these are young, highly star-forming galaxies with significant amounts of dust extinction. However, when the UV photometry is included, the best-fit SEDs suggest that these galaxies are on average more massive, significantly older and less dusty and with significantly lower star formation rates than inferred from SED fits without the UV data. In Figure \ref{fig:nir}, these galaxies migrate from the top left portion of the top panel, towards the bottom right portion of the middle panel. The UV photometry is dominated by light from instantaneous star formation rather than reflecting the total underlying stellar mass. In most galaxies, it therefore provides little constraint on the total stellar mass. However, we see that for some of the WiggleZ galaxies, the lack of UV photometry can lead to an over-estimate of the dust extinction and star formation rate and an under-estimate of the age of the galaxy. As a consequence of the SEDs being fit by younger stellar populations, the total stellar mass is also under-estimated without the UV data, in these galaxies. 

We also consider the effect of the UV photometry on the 27,305 fainter, bluer galaxies that are undetected in the IR and therefore where NIR photometry is no longer included in the SED fitting. Once again, there is little effect on the median stellar mass, which goes down by 0.01 dex on removing the \textit{GALEX} bands from the fits, and the median 1$\sigma$ errors remain unchanged. Almost 90\% of galaxies now have stellar masses that are consistent within the 1$\sigma$ errors when considering the outputs with and without the UV photometry. These 27,305 galaxies have no NIR data and in the case of these galaxies, we find no evidence for a separate cloud of galaxies where the lack of UV photometry leads to poorer constraints. 

\subsection{Effect of Stellar Population Synthesis Models}

\label{sec:sps}

We now consider the effect of different SPS models on the stellar mass estimates with all other parameters in the SED fitting held constant. We use a Salpeter IMF and a set of simple single component exponentially decaying SFHs. FAST allows fitting of both the BC03 and \citet{Maraston:05} models to the multi-wavelength data. While the latter includes a prescription for TP-AGB stars, the BC03 models do not. The contribution of TP-AGB stars to the NIR colours of galaxies is still widely debated with local galaxies showing significant dependences of the colours on this component \citep{Eminian:08} while the contribution in higher redshift galaxies is deemed to be less significant \citep{Kriek:10}. TP-AGB stars are only expected to affect the NIR colours of intermediate age ($\sim$1Gyr) stellar populations. We also derive stellar masses using the KG04 code and a set of PEGASE.2 SEDs, in which nebular emission lines can be turned on or off in order to consider their effect on the stellar masses. Note that although the KG04 code incorporates the facility to include secondary bursts of star formation, for this particular test, only single component SFHs are allowed so as to allow direct comparison with the results from FAST where secondary bursts cannot be included. Having already demonstrated that the NIR data can help in better constraining the stellar masses for FAST, and that it does not materially change the stellar masses for the KG04 code, we use the UV, optical and NIR photometry in the SED fits in all cases where the NIR is available. We keep the SFH and IMF fixed so the effect of the different models on the stellar mass estimates can be isolated. 

\begin{figure*}
\begin{center}
\begin{tabular}{ccc}
\includegraphics[scale=0.35]{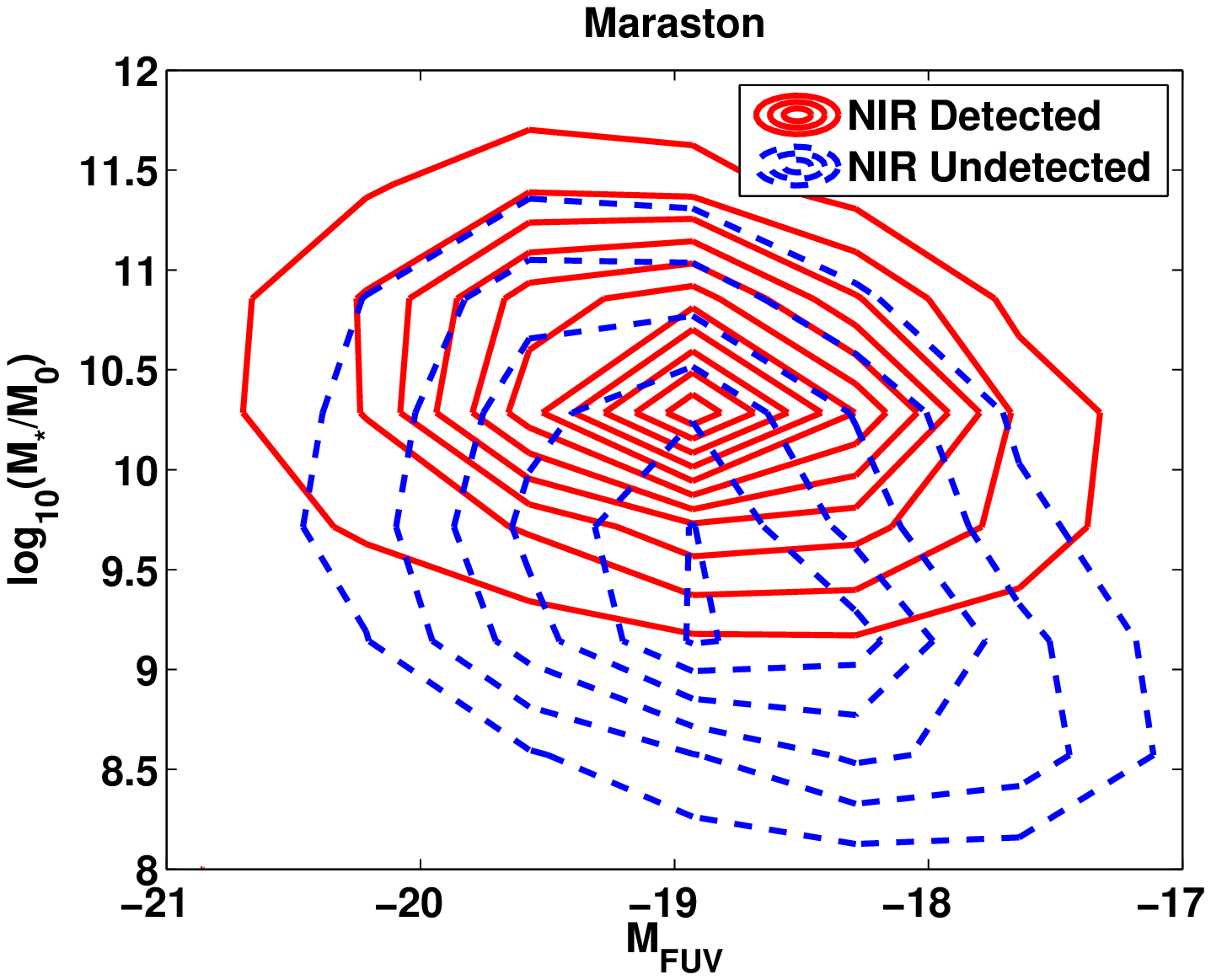} & \includegraphics[scale=0.35]{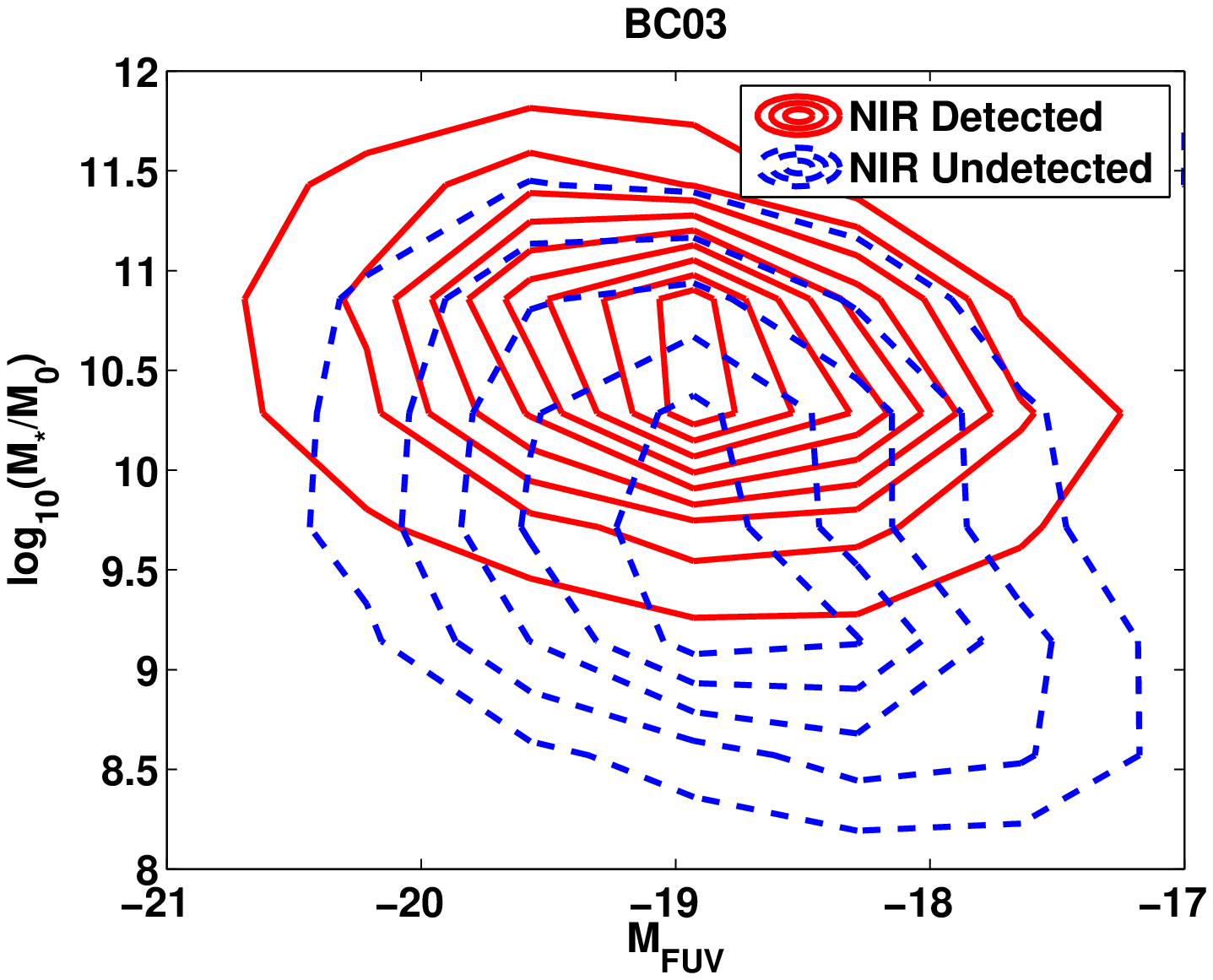} & \includegraphics[scale=0.35]{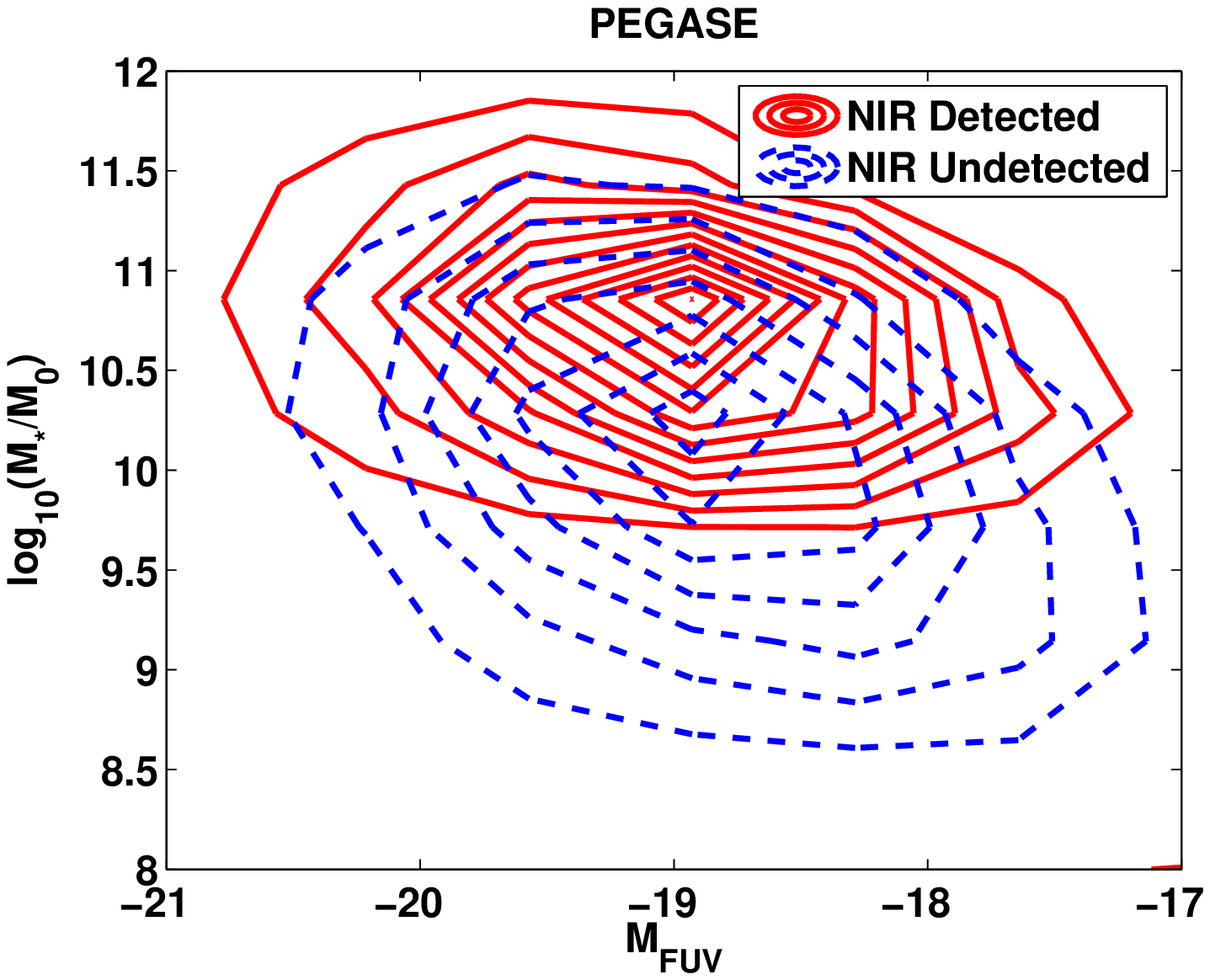} \\
\end{tabular}
\caption{Stellar mass versus M$_{FUV}$ for the WiggleZ galaxies undetected in the IR (blue dashed) and those detected in the IR (red solid) for three different choices of SPS models. Contours represent the density of points computed using a kernel density estimator, over a grid traversing the parameter space shown. The stellar masses are lowest for the Maraston models which include TP-AGB stars, followed by BC03 and then PEGASE, which produces the largest masses. The discrepancy between PEGASE and BC03/Maraston, is largest at faint M$_{FUV}$ and particularly noticeable for the blue cloud of IR undetected galaxies. The discrepancy between BC03 and Maraston on the other hand affects more the IR detected, bright M$_{FUV}$ galaxies that populate the top left corner of the panels.}
\label{fig:model_mass}
\end{center}
\end{figure*}

For the NIR detected population, we find the reduced $\chi^2$ values and 1$\sigma$ errors on the stellar masses to be very similar for the BC03 and Maraston models. However, the median stellar mass is $\sim$0.2 dex lower when using the Maraston models. For the bluer subset of galaxies undetected in the NIR, the average difference between the Maraston and BC03 derived stellar masses now falls to $\sim$0.1 dex. Figure \ref{fig:model_mass} shows the distribution of stellar mass as a function of M$_{FUV}$ for both WiggleZ galaxies detected and undetected in the NIR for each of the different SPS models studied. Changing from the BC03 models to the Maraston models primarily affects the red cloud of galaxies detected in the IR. This is expected given that we infer older ages for these IR detected galaxies suggesting that they have slightly more evolved stellar populations where TP-AGB stars may become more important. 

Before considering the differences between the BC03/Maraston and PEGASE.2 models, we first assess the effect that nebular emission lines have on the stellar masses, by computing these masses with the nebular emission switched on and off in the PEGASE models. We find that turning off the nebular emission in the models increases the stellar masses as expected by raising the best-fit stellar continuum. However, the median increase in the stellar masses due to lack of nebular emission in the models, is only $\sim$0.01/0.02 dex for those galaxies detected in the IR and those that are undetected respectively. The WiggleZ galaxies are selected to have strong emission lines in order to aid the acquisition of redshifts for cosmology \citep{Drinkwater:10}. However, even in these strong emitters, the impact of including nebular emission in the models on the stellar masses, appears to be negligible for the majority of the galaxies and smaller than the offsets typically seen in high redshift LBGs (e.g. \citealt{deBarros:12,Stark:12}). We note however, that the prescription for nebular emission in the PEGASE.2 SEDs, is relatively simple and does not for example allow for variations in the line ratios. There are of course WiggleZ galaxies where the emission lines make a significant contribution to the broadband flux, and where the inclusion of nebular emission in the models, results in a more significant offset to the stellar mass. Some example SED fits of galaxies with strong nebular emission can be found in Appendix A. However, on average, the effect of including emission lines in the models, on the stellar masses, is found to be very small.

Having demonstrated that the inclusion of nebular emission lines in the SPS models, appears to have little impact on the median stellar masses of the WiggleZ galaxies, we now look at the differences in stellar mass between the PEGASE.2 models and the Maraston and BC03 models. The results can be seen in Figure \ref{fig:model_mass}. We find that the PEGASE.2 models always produce stellar masses that are higher than those derived by the BC03 and Maraston models. These masses are higher by $\sim$0.1 (0.2) dex with respect to BC03 (Maraston) for the galaxies detected in the IR, and $\sim$0.2 (0.3) dex with respect to BC03 (Maraston) for the bluer galaxies undetected in the IR. We have ruled out nebular emission as the cause for these mass differences, and shown that the presence of emission lines would decrease rather than increase the stellar masses computed with the PEGASE models. These differences, which seem to be accentuated at faint $FUV$ luminosities, must therefore result from differences in the input stellar libraries in each of the models and the fact that these stellar libraries traverse different regions in physical parameter space. We note that restricting the PEGASE models to solar metallicity only, produces slightly more consistent stellar masses between PEGASE and BC03 by reducing the PEGASE produced stellar masses by $\sim$0.1 dex.

Unlike the stellar masses which are reasonably well constrained, the best-fit ages are once again more poorly constrained and show significant deviations between the models. Nebular emission is found to produce younger ages and the PEGASE.2 models generally produce higher median ages than BC03 and Maraston. However, we emphasise that the uncertainties on these age estimates are considerable. 

\subsection{Effect of Different Initial Mass Functions}

\label{sec:imf}


The form of the stellar initial mass function assumed during SED fitting, can have a significant effect on the inferred stellar masses of galaxies. In this section, we systematically quantify the difference in stellar masses using different IMFs but assuming the same SPS models, photometric filters and star formation histories. We use the subset of 11,919 WiggleZ galaxies matched to UKIDSS, for the comparison. We change the IMF choice in both FAST and KG04 while keeping the rest of the input parameters fixed, in order to quantify the resulting change in the median stellar masses of the WiggleZ galaxies. 

Firstly, using FAST with the BC03 models and exponentially decaying SFHs, we find that the median difference between the Salpeter and Chabrier IMFs is $\sim$0.24 dex. The Chabrier IMF has the same power-law slope as Salpeter at the high-mass end but turns over at M$<$1M$_\odot$. 

FAST is also used with the Maraston models and exponentially decaying SFHs to look at differences in stellar mass between the Salpeter and \citet{Kroupa:01} IMF. The Kroupa IMF predicts stellar masses that are on average $\sim$0.20 dex lower than Salpeter. It too has the same power-law slope as Salpeter at the high-mass end but turns over at M$<0.8$M$_\odot$. These differences in stellar mass due to the IMF, show little dependence on the $FUV$ luminosity or galaxy colour. 

Finally, we use the KG04 code with PEGASE models and multi-component SFHs, to look at the difference between the Salpeter, Kroupa, BG03 and Chabrier IMFs and to check if these differences are enhanced or reduced when additional bursts of star formation are allowed. The BG03 IMF has a shallower slope than Salpeter at the high-mass end and turns over at M$<$0.5M$_\odot$. It, results in stellar masses that are on average 0.24 dex lower than Salpeter \citep{Glazebrook:04}. The stellar masses obtained with the BG03 and Chabrier IMFs are remarkably similar with a median difference of $<$0.001 dex between them. The Kroupa IMF results in stellar masses that are $\sim$0.1 dex higher than those produced by BG03 and Chabrier. 

The simple power-law form of the Salpeter IMF is known to overestimate the number of low-mass stars in galaxies and therefore the stellar mass \citep{Cappellari:06, Ferreras:08} and more recent analytical forms of the IMF such as those considered in this work, have, by design a more realistic break in the IMF at low masses and therefore predict lower stellar masses on average.  We find that the Kroupa IMF results in stellar masses that are $\sim$0.1--0.2 dex lower than Salpeter while the BG03 and Chabrier IMFs produce very similar stellar masses that are $\sim$0.24 dex lower than Salpeter and up to 0.1 dex lower than Kroupa.   

 
\subsection{Effect of Different Star Formation Histories}

\subsubsection{Single Component Star Formation Histories}

In order to study the dependence of the stellar mass estimates on the assumed SFH, we begin by using FAST to fit BC03 models with three different star formation histories to the photometric data. These correspond to a simple exponentially decaying SFH, a truncated SFH which corresponds to constant star formation between t$_{\rm{onset}}$ and t$_{\rm{onset}}$+$\tau$, and a delayed exponentially decaying star formation history with SFR$\sim$ t exp(-t/$\tau$). We note that the FAST code used for the stellar mass estimates currently does not incorporate the ability to include exponentially increasing or inverted-$\tau$ models which have recently been argued to provide a better description of the extinction and star formation rates inferred from rest-frame UV data alone for galaxies at z$>$1 \citep{Maraston:10}.  

\begin{figure}
\begin{center}
\includegraphics[width=8.5cm,height=6.0cm,angle=0] {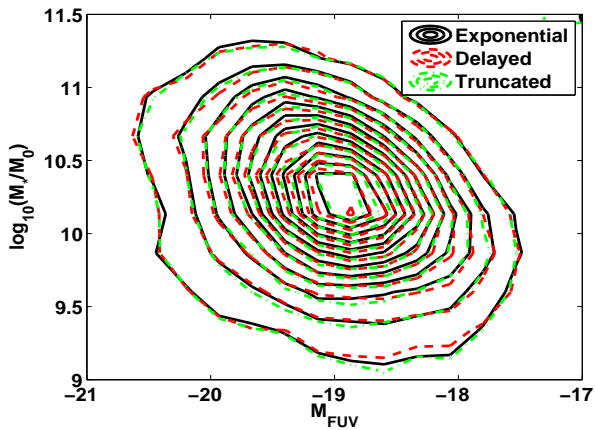}
\caption{Stellar mass distribution as a function of M$_{FUV}$, for the IR detected WiggleZ galaxies, computed with various different prescriptions for single component SFHs. The contours represent the density of points estimated using a kernel density estimator over a grid covering the parameter space shown. The stellar mass is seen to be insensitive to the choice of star formation history for single component SFHs.}
\label{fig:sfh_mass}
\end{center}
\end{figure}

In Figure \ref{fig:sfh_mass} we show the distribution of stellar masses for the 11,919 UKIDSS detected WiggleZ galaxies as a function of M$_{FUV}$ for each of the three SFHs. We find that with a large sample of spectroscopically confirmed galaxies such as ours, the stellar mass is insensitive to the choice of star formation history for simple single component SFHs. However, as might be expected for these very blue emission-line galaxies, the truncated SFH produces relatively poor reduced-$\chi^2$ values as it is typically invoked to describe the SFHs of more passive systems such as quiescent and post-starburst galaxies \citep{Daddi:04, Kriek:10}. 

Although the stellar masses are insensitive to the choice of SFH, the ages and dust extinctions, which are in general more poorly constrained, show significant variations as the SFH is changed and are not particularly well constrained. 


\subsubsection{Addition of Bursts}

\label{sec:bursts}

Although the stellar mass is found to be insensitive to the assumed star formation history for a single component SFH, it is interesting to quantify the effect of more complex SFHs on the mass estimates. FAST does not allow for addition of random bursts of star formation on top of the single component star formation history. In order to assess the effect of these bursts on the stellar masses, we use the KG04 code with the PEGASE.2 SEDs including nebular emission and assuming a BG03 IMF. As demonstrated in Section \ref{sec:imf}, the choice of IMF has little effect on the stellar masses for IMFs with realistic breaks at low masses.  We examine the impact of starbursts on the stellar masses, both for the IR detected subset of 11,919 galaxies as well as the bluer 27,305 galaxies that are undetected in wide-field IR surveys. 

\begin{figure*}
\begin{center}
\begin{minipage}[c]{1.00\textwidth}
\centering
\includegraphics[scale=0.7] {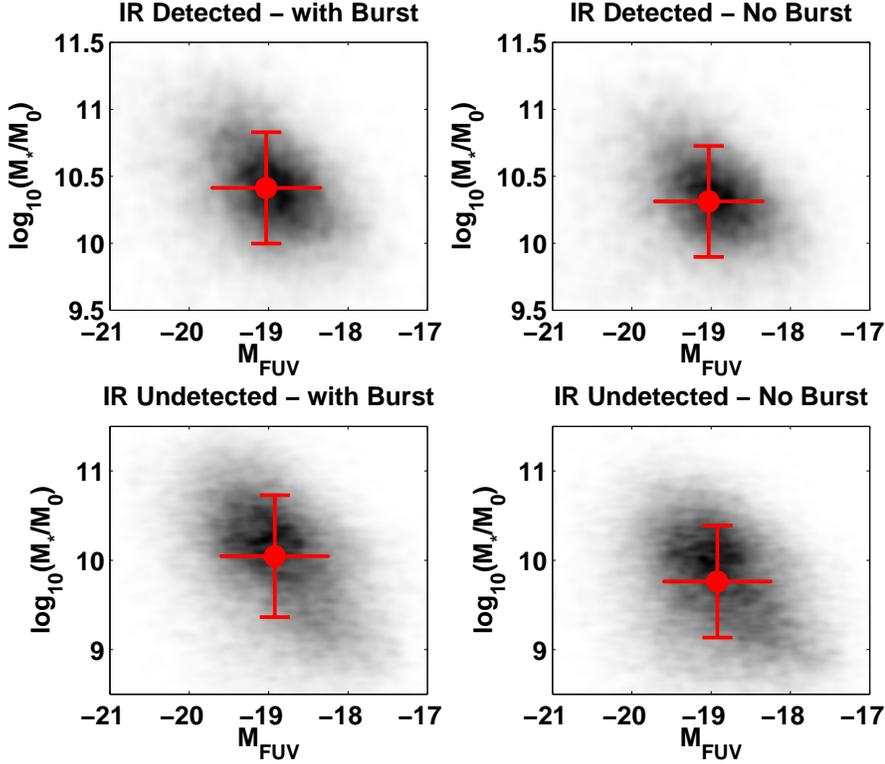}
\end{minipage}
\caption{Stellar mass distribution as a function of $FUV$ absolute magnitude for WiggleZ galaxies detected and undetected in infrared surveys. The greyscale denotes the density of points while the individual points with error bars represent the median and standard deviations of the samples. Stellar masses are calculated using the KG04 code, PEGASE SEDs, assuming a BG03 IMF and with additional bursts of star formation turned on or off. Additional bursts of star formation add $\sim$0.1 dex in stellar mass to galaxies detected in the NIR and $\sim$0.3 dex of stellar mass in galaxies undetected in the NIR. The effect is more pronounced at brighter $FUV$ luminosities. }
\label{fig:burst}
\end{center}
\end{figure*}

The results are shown in Figure \ref{fig:burst} where we plot the stellar masses as a function of M$_{FUV}$ for both subsets of galaxies with and without additional bursts of star formation. We find that the addition of bursts of star formation leads to an increase in the stellar masses in both cases. This increase is more pronounced for those galaxies not detected in the IR where the addition of bursts adds $\sim$0.3 dex to the stellar masses, versus 0.1 dex for galaxies detected in the IR. For both subsamples, the increase in stellar mass on inclusion of bursts is also more pronounced at higher $FUV$ luminosities. This is because the secondary bursts essentially ``hide'' the more massive evolved stellar populations by dominating the UV-light. 

We check these results using the MAGPHYS code, which makes different assumptions in the implementation of the secondary burst as discussed in detail in Section \ref{sec:sed}. As previously stated, MAGPHYS is not readily applicable to the large sample sizes assembled here, but, unlike FAST, it does include additional bursts of star formation. We therefore compare the FAST and MAGPHYS derived stellar masses for a randomly selected sub-sample of 800 WiggleZ galaxies that are also detected in UKIDSS and a similar sub-sample of WiggleZ galaxies not detected in UKIDSS. We find that the mean difference in stellar mass between FAST and MAGPHYS is 0.1 dex for the UKIDSS detected galaxies and 0.25 dex for the UKIDSS undetected galaxies, consistent with the results derived using the KG04 code due to the effect of bursts. 

\begin{figure}
\begin{center}
\includegraphics[width=8.5cm,height=6.0cm,angle=0] {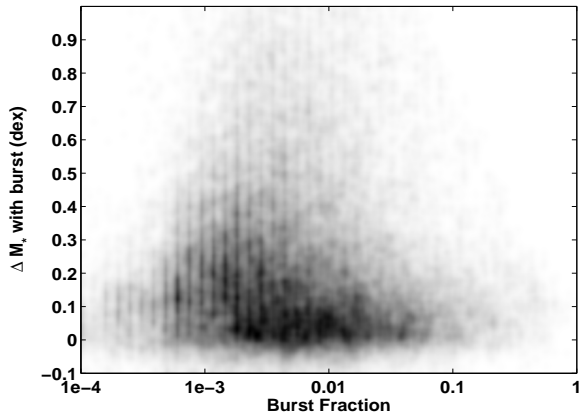}
\caption{Difference in stellar mass obtained using the KG04 code and PEGASE models on inclusion of bursts, versus the burst mass fraction for all 39,701 WiggleZ galaxies. The greyscale denotes the density of points. The median burst fraction is $\sim$1 per cent and the inclusion of the bursts in the SED models, adds $\sim$0.1--0.3 dex of stellar mass on average to the galaxies.}
\label{fig:fburst}
\end{center}
\end{figure}

In Figure \ref{fig:fburst}, we plot the difference in stellar mass on inclusion of bursts versus the burst mass fraction derived using KG04 for all $\sim$40,000 WiggleZ galaxies. Most of the galaxies have typical burst fractions of $\sim$1\% - i.e. the mass produced in the secondary burst is just 1\% of that produced through continuous star formation. However Figure \ref{fig:fburst} clearly shows that the difference in stellar mass between the burst and no burst SED fits, is not particularly well correlated with the burst fraction. This demonstrates that allowing secondary bursts in the SED fits, does not just change the total stellar mass of the galaxy but also results in best-fit SEDs with different properties - age, $A_V$, $\tau$ - compared to when bursts are not included in the SED fits. 

For multi-component SFHs, the ages output are the mass-weighted ages of the different stellar populations so a direct comparison to the ages produced by fitting single component SFHs, cannot be made. The median mass weighted ages are $\sim$2--3 Gyr. The addition of secondary bursts therefore produces older galaxies than when these bursts are not included. This is because the underlying primary stellar population being hidden by the current burst is also older than when the bursts are not included. 

\begin{figure*}
\begin{center}
\begin{tabular}{cc}
\includegraphics[scale=0.5,angle=0] {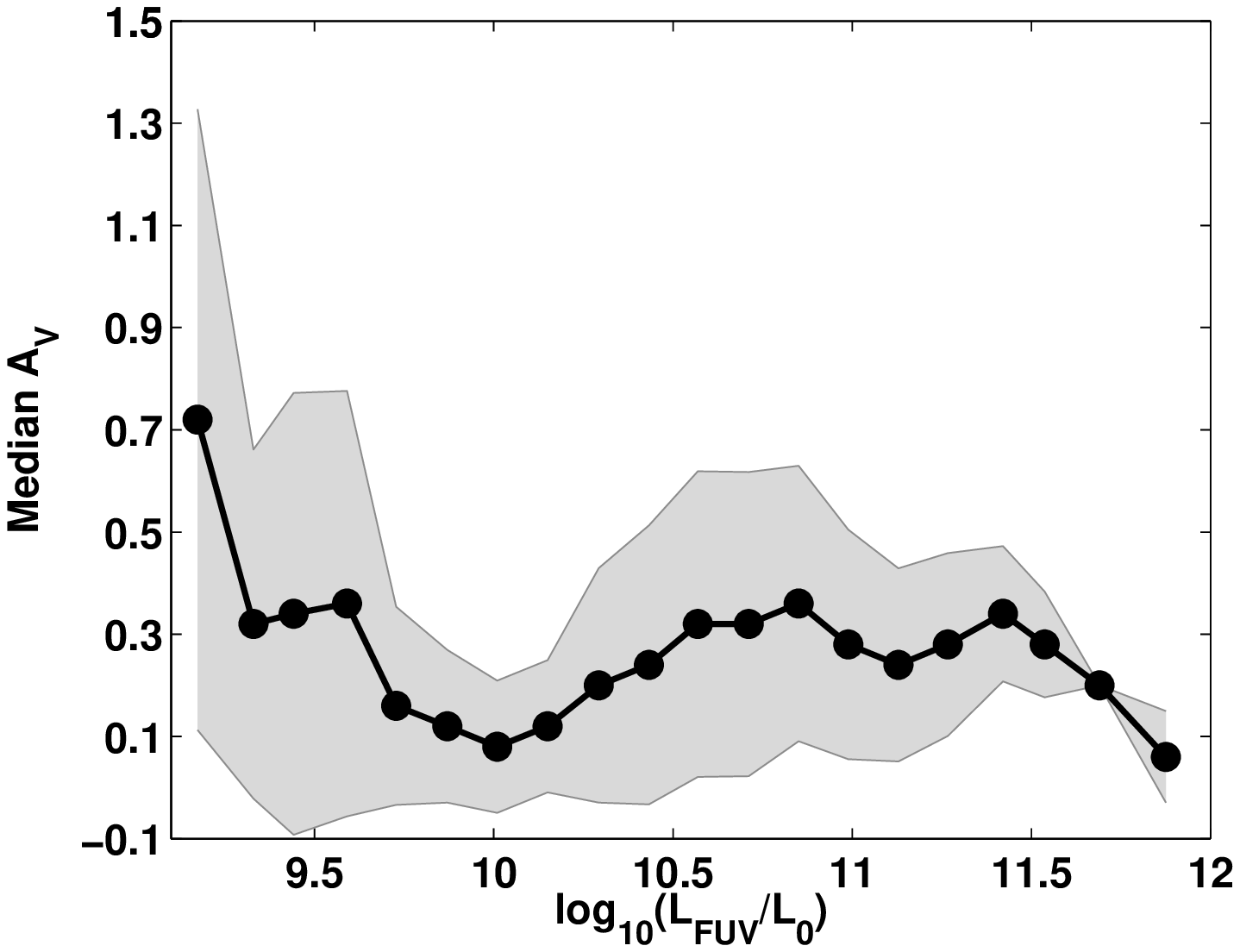} & \includegraphics[scale=0.5,angle=0] {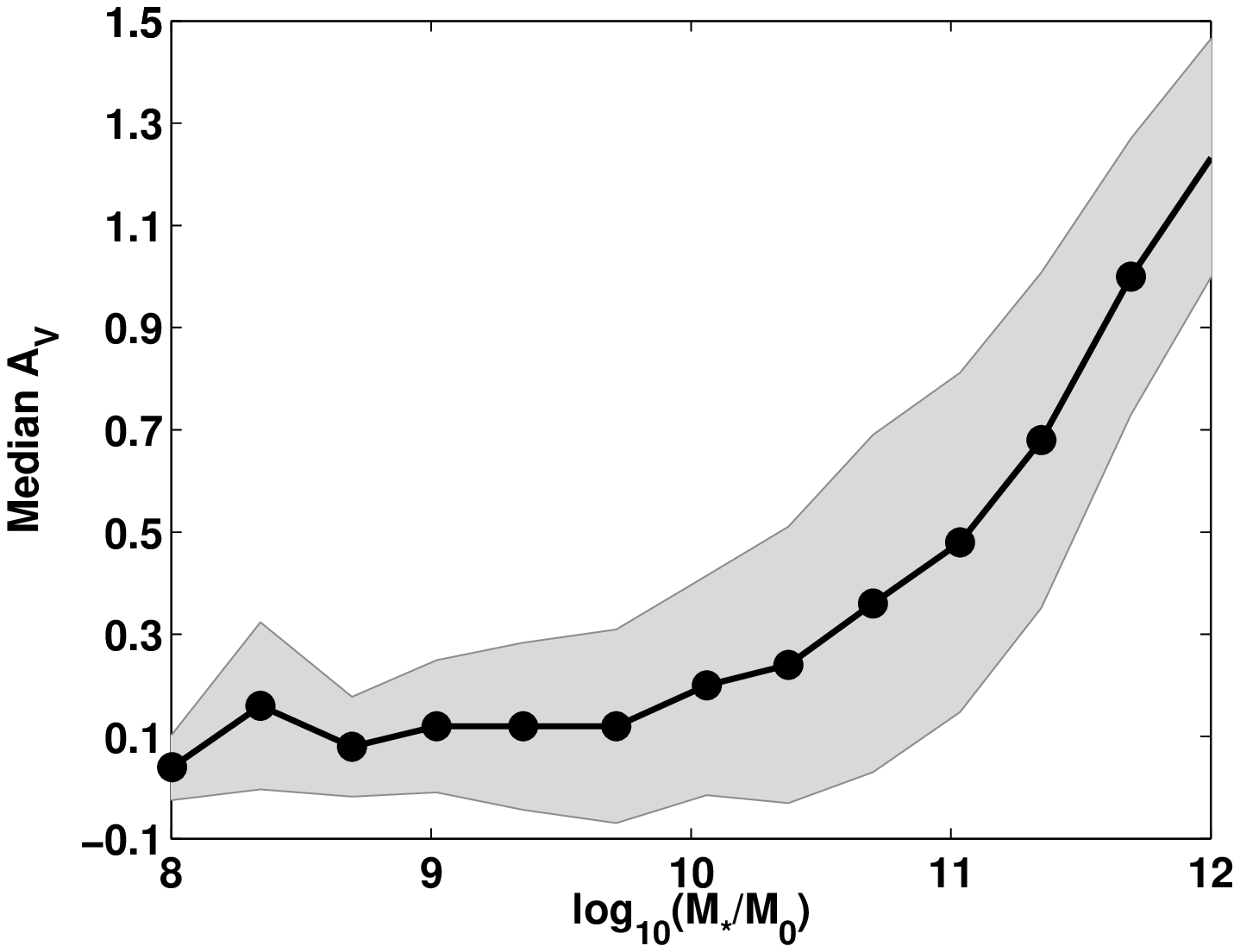} \\
\end{tabular}
\caption{Median best-fit dust extinction parameter A$_V$ derived using the KG04 code with the PEGASE.2 SPS models, as a function of $FUV$ luminosity and stellar mass for all $\sim$40,000 WiggleZ galaxies. The points represent the median values in bins of L$_{FUV}$ and stellar mass with the shaded regions denoting the 1$\sigma$ errors on A$_V$ from the SED fitting. Despite the large errors, we note a trend of increasing dust extinction with increasing stellar mass and decreasing L$_{FUV}$ in this sample.}
\label{fig:Av}
\end{center}
\end{figure*}

We also consider the variation of the best-fit dust extinction parameter $A_V$ with $FUV$ luminosity and stellar mass. We consider the outputs from the KG04 code including both nebular emission as well as secondary bursts of star formation. Although these dust extinctions are typically not very well constrained with median 1$\sigma$ errors of $\sim$0.2 mags, we can nevertheless assess whether there is any evidence for a trend in best fit A$_V$ with the $FUV$ luminosity or stellar mass. Figure \ref{fig:Av} shows the median best-fit A$_V$ in bins of L$_{FUV}$ and stellar mass for all $\sim$40,000 WiggleZ galaxies with the shaded regions representing the 1$\sigma$ errors from the SED fits. Despite the large errors, we find a strong trend with stellar mass with the most massive galaxies also being fit to have larger dust extinction values. The trend with $FUV$ luminosity is less strong but there is some evidence for WiggleZ galaxies with lower $FUV$ luminosities also having higher best-fit $A_V$. These trends are confirmed from the outputs derived using the FAST code which includes only single component SFHs and no nebular emission lines in the SPS models. Recently, \citet{Buat:12} observed a similar trend in dust attenuation using \textit{Herschel} data to analyse a sample of intermediate redshift galaxies in the GOODS field. \citet{Heinis:12} too find a slight increase in dust attenuation at low $FUV$ luminosities although the trend is essentially flat at L$_{FUV}>10^{10}$L$_\odot$, once again consistent with our data in Figure \ref{fig:Av}. However, given the biased colour selection of WiggleZ targets and the considerable uncertainties and degeneracies involved in accurately constraining A$_V$ from SED fits to the available broadband photometry for this sample, we caution against over interpretation of these observed trends.

\subsection{Summary \& $K$-Band Mass-to-Light Ratios}

We have quantified the sensitivity of stellar mass estimates for the very blue population of WiggleZ galaxies at $0.3<z<1.0$  to changes in various input parameters and assumptions made during the SED fitting. In particular, we have looked at how the median stellar masses of this large sample are affected by changes to the input photometry, choice of SPS model, choice of IMF and choice of star formation history. The SPS models and the addition of secondary bursts result in the biggest changes to the median stellar masses - of the order of 0.3 dex for these very blue galaxies. Table \ref{tab:mass_summary} summarises the results of the various SED fits carried out in this Section and provides median stellar masses for different subsamples of the WiggleZ population output using different codes and different input parameters to the SED fitting. A key result of this paper is that the stellar masses seem to be relatively robust to changes in the various input parameters to the SED fitting as well as the choice of SED fitting code. 

\onecolumn
\begin{landscape}
\small
\begin{table*}
\caption{Summary of median stellar masses of WiggleZ galaxies computed using different SED fitting codes and when changing different input parameters in the SED fitting.}
\label{tab:mass_summary}
\begin{center}
\begin{tabular}{cccccccc}
\hline
Code & N$_{\rm{gal}}$ & Filters & SPS Model & Nebular Lines & SFH & IMF & $<$log$_{10}$(M$_*$/M$_\odot$)$>$ \\
\hline \hline
FAST & 27,305 & GALEX,$ugriz$ & BC03 & No & Exp decaying & Chabrier & 9.6$\pm$0.7 \\
FAST & 27,305 & $ugriz$ & BC03 & No & Exp decaying & Chabrier & 9.5$\pm$0.6 \\
FAST & 11,919 & GALEX,$ugrizYJHK$ & BC03 & No & Exp decaying & Chabrier & 10.2$\pm$0.5 \\
FAST & 11,919 & $ugrizYJHK$ & BC03 & No & Exp decaying & Chabrier & 10.2$\pm$0.4 \\
FAST & 11,919 & GALEX,$ugriz$ & BC03 & No & Exp decaying & Chabrier & 10.4$\pm$0.6 \\
FAST & 6117 & GALEX,$ugrizYJHK$,3.4,4.6$\mu$m & BC03 & No & Exp decaying & Chabrier & 10.4$\pm$0.5 \\
FAST & 11,919 & GALEX,$ugrizYJHK$ & BC03 & No & Exp decaying & Salpeter & 10.4$\pm$0.6 \\
FAST & 11,919 & GALEX,$ugrizYJHK$ & Maraston & No & Exp decaying & Salpeter & 10.3$\pm$0.6 \\
KG04 & 11,919 & GALEX,$ugrizYJHK$ & PEGASE & Yes & Exp decaying & Salpeter & 10.5$\pm$0.4 \\
KG04 & 11,919 & GALEX,$ugriz$ & PEGASE & Yes & Exp decaying & Salpeter & 10.5$\pm$0.4 \\
KG04 & 11,919 & GALEX,$ugrizYJHK$ & PEGASE & No & Exp decaying & Salpeter & 10.5$\pm$0.4 \\
KG04 & 27,305 & GALEX,$ugriz$ & PEGASE & Yes & Exp decaying & Salpeter & 10.0$\pm$0.6 \\
KG04 & 27,305 & GALEX,$ugriz$ & PEGASE & No & Exp decaying & Salpeter & 10.0$\pm$0.6 \\
FAST & 27,305 & GALEX,$ugriz$ & BC03 & No & Exp decaying & Salpeter & 9.8$\pm$0.7 \\
FAST & 27,305 & GALEX,$ugriz$ & Maraston & No & Exp decaying & Salpeter & 9.7$\pm$0.7 \\
KG04 & 27,305 & GALEX,$ugriz$ & PEGASE & No & Exp decaying & Salpeter & 10.0$\pm$0.6 \\
FAST & 11,919 & GALEX,$ugrizYJHK$ & BC03 & No & Delayed Exp & Chabrier & 10.2$\pm$0.5 \\
FAST & 11,919 & GALEX,$ugrizYJHK$ & BC03 & No & Truncated & Chabrier & 10.2$\pm$0.5 \\
KG04 & 11,919 & GALEX,$ugrizYJHK$ & PEGASE & Yes & Exp decaying + Burst & BG03 & 10.4$\pm$0.4 \\
KG04 & 27,305 & GALEX,$ugriz$ & PEGASE & Yes & Exp decaying + Burst & BG03 & 10.0$\pm$0.7 \\
KG04 & 11,919 & GALEX,$ugrizYJHK$ & PEGASE & Yes & Exp decaying + Burst & Kroupa & 10.5$\pm$0.4 \\
KG04 & 11,919 & GALEX,$ugrizYJHK$ & PEGASE & Yes & Exp decaying + Burst & Chabrier & 10.4$\pm$0.4 \\
KG04 & 11,919 & GALEX,$ugrizYJHK$ & PEGASE & Yes & Exp decaying + Burst & Salpeter & 10.7$\pm$0.4 \\
FAST & 11,919 & GALEX,$ugrizYJHK$ & Maraston & No & Exp decaying & Kroupa & 10.0$\pm$0.5 \\
MAGPHYS & 24 & GALEX,$ugrizYJHK$,3.4,4.6,12,22$\mu$m & CB07 & No & Exp decaying + Burst & Chabrier & 10.8$\pm$0.3 \\
MAGPHYS & 800 & GALEX,$ugrizYJHK$ & CB07 & No & Exp decaying + Burst & Chabrier & 10.4$\pm$0.5 \\
MAGPHYS & 800 & GALEX,$ugriz$ & CB07 & No & Exp decaying + Burst & Chabrier & 9.9$\pm$0.7 \\
\hline
\end{tabular}
\end{center}
\end{table*}
\twocolumn
\end{landscape}

We have also demonstrated that the near infra-red data can, for some codes and choice of priors, significantly reduce the 1$\sigma$ errors on the stellar masses by tightening the upper bound on the stellar mass estimate. We find no evidence that the near infra-red data actually worsens the quality of the SED fits for the WiggleZ galaxies with any of the codes. As previously mentioned, rest-frame near infra-red mass-to-light ratios have been used for some time to estimate the total stellar masses of galaxies. These mass-to-light ratios are significantly less dependent on galaxy colour than similar ratios derived at shorter wavelengths. Empirical relations of the dependence of stellar mass-to-light ratios on galaxy colour, have been derived by \citet{Bell:03} and are widely used in the literature. We therefore now compare this empirical relation in the near infrared $K$-band to the equivalent mass-to-light ratios derived from our SED fits for our sample of almost 12,000 WiggleZ galaxies that are detected in UKIDSS. Note that at the median redshift of our sample, the rest-frame $K$-band is not sampled by the UKIDSS data. However, the availability of the UKIDSS photometry allows us to sample more of the galaxy SED than is sampled by the optical, making the extrapolation to the rest-frame $K$-band more secure. We consider the outputs from the KG04 code including secondary bursts of star formation and nebular emission lines in the SPS models.  

\begin{figure}
\begin{center}
\includegraphics[width=8.5cm,height=6.0cm,angle=0] {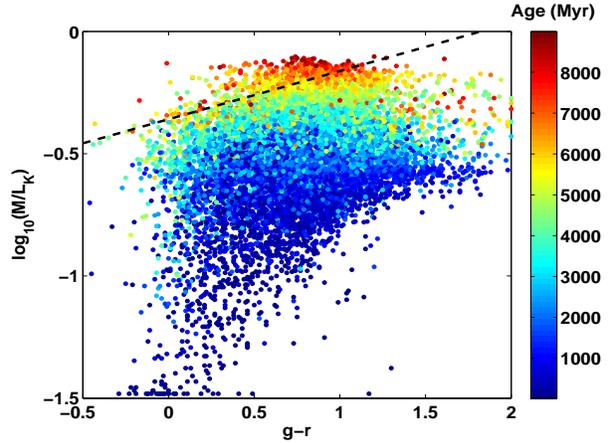}
\caption{Stellar mass-to-light ratio in the rest-frame $K$-band as a function of the observed $(g-r)$ colour, derived from SED fits using the KG04 code and including both secondary bursts of star formation and nebular emission in the models. The points are colour-coded according to the best-fit age of the galaxy. The dashed line shows the empirical colour-based estimator of the M/L$_K$ from \citet{Bell:03}. This simple colour-based estimator over-predicts the mass-to-light ratio, particularly for younger, bluer galaxies.}
\label{fig:MLK}
\end{center}
\end{figure}

Figure \ref{fig:MLK} shows the rest-frame $K$-band mass-to-light ratio computed from the best-fit SEDs, as a function of the $(g-r)$ colour for those WiggleZ galaxies with UKIDSS NIR photometry. Note, that the rest-frame $K$-band luminosity is calculated here using the best-fit SED for each galaxy and not the LBG template used to calculate the rest-frame $FUV$ luminosity. The rest-frame $K$-band luminosity depends more critically on the exact choice of SED used to derive it. The \citet{Bell:03} colour-based estimator is also shown with an appropriate offset to match the IMF used in our SED fits. Although these mass-to-light ratios are computed using the KG04 code which does not include a prescription for dust emission in the NIR, we find in Table \ref{tab:mass_summary}, that the KG04 derived stellar masses and MAGPHYS derived stellar masses are very similar. As MAGPHYS does include a prescription for dust emission in the NIR, we conclude that this emission has negligible impact on our calculated stellar masses and mass-to-light ratios. We find that even with the inclusion of secondary bursts, the stellar mass-to-light ratios derived from the SEDs, are lower than predicted by the colour-based estimator. The simple colour-based estimator over predicts the mass-to-light ratio by $\sim$0.4 dex on average compared to the more sophisticated SED fitting approach in the case of these WiggleZ galaxies. Almost 75\% of the WiggleZ galaxies have $K$-band mass-to-light ratios that are formally inconsistent with the simple colour-based estimator after taking into account the errors on M/L$_K$ from the SED fitting. As seen in Figure \ref{fig:MLK}, this is particularly true for the younger, bluer galaxies. For the oldest and reddest WiggleZ galaxies with ages $\gtrsim$7Gyr, the SED derived M/L$_K$ and the colour-based estimator, agree reasonably well. As expected, the WiggleZ colour cuts select very few massive red galaxies that would populate the top right corner of Figure \ref{fig:MLK}. 

\section{MID-INFRARED LUMINOUS WIGGLEZ GALAXIES}

\label{sec:ir}

\begin{figure}
\begin{center}
\includegraphics[width=8.5cm,height=6.0cm,angle=0] {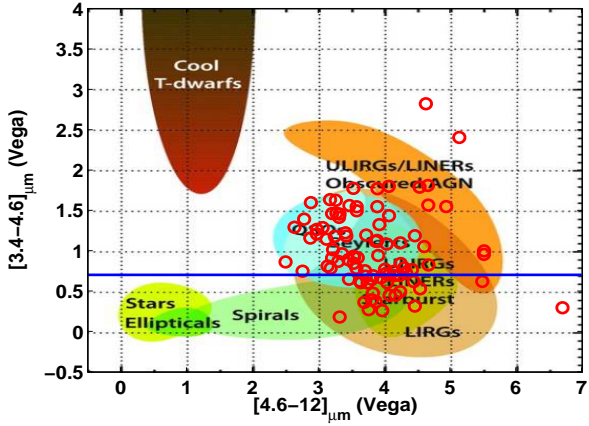}
\caption{The location of our WiggleZ mid-IR bright galaxies in the [3.4$-$4.6]$_{\mu\rm{m}}$ versus [4.6$-$12]$_{\mu\rm{m}}$ colour-colour plane of \citep{Wright:10}. The WiggleZ galaxies are shown as the red circles. While most are found to be mid-IR bright on account of some AGN contamination, we isolate a population of 24 galaxies with [3.4$-$4.6]$_{\mu\rm{m}}<0.7$ that lie below the solid horizontal line in the figure and can reasonably be assumed to be star-formation dominated.}
\label{fig:wise_colour}
\end{center}
\end{figure}

Before discussing the results of our SED fitting and stellar mass estimates within the broader context of galaxy evolution, we briefly digress to look at the properties of the very small subset of WiggleZ galaxies that are found to be extremely luminous at 12 and 22$\mu$m. There are 78 galaxies at 0.3$<z<1.0$ out of the total sample of 39,701, that are detected at $>$5$\sigma$ in \textit{WISE} at 12 and 22$\mu$m. In Figure \ref{fig:wise_colour}, we plot their colours in the [3.4$-$4.6]$_{\mu\rm{m}}$ versus [4.6$-$12]$_{\mu\rm{m}}$ plane, taken from figure 12 of \citet{Wright:10}. We find unsurprisingly, that the mid infrared emission can be accounted for by the presence of an AGN in the majority of these galaxies with the \textit{WISE} colours of the galaxies overlapping the QSO and Obscured AGN loci of \citet{Wright:10}. In order to select those galaxies where the AGN contamination should be less significant, we apply a conservative cut of [3.4$-$4.6]$_{\mu\rm{m}}<0.7$ \citep{Stern:12}, which is also shown in Figure \ref{fig:wise_colour}. This leaves us with only 24 galaxies which may reasonably be assumed to be star-forming and lie in the starburst/LIRG regime of the colour-colour plane. We visually inspect the \textit{WISE} images for these 24 galaxies and eliminate two which lie in the halos of bright sources in the \textit{WISE} data.

\begin{figure}
\begin{center}
\begin{tabular}{c}
\vspace{-1.5cm}
\includegraphics[scale=0.5,angle=0]{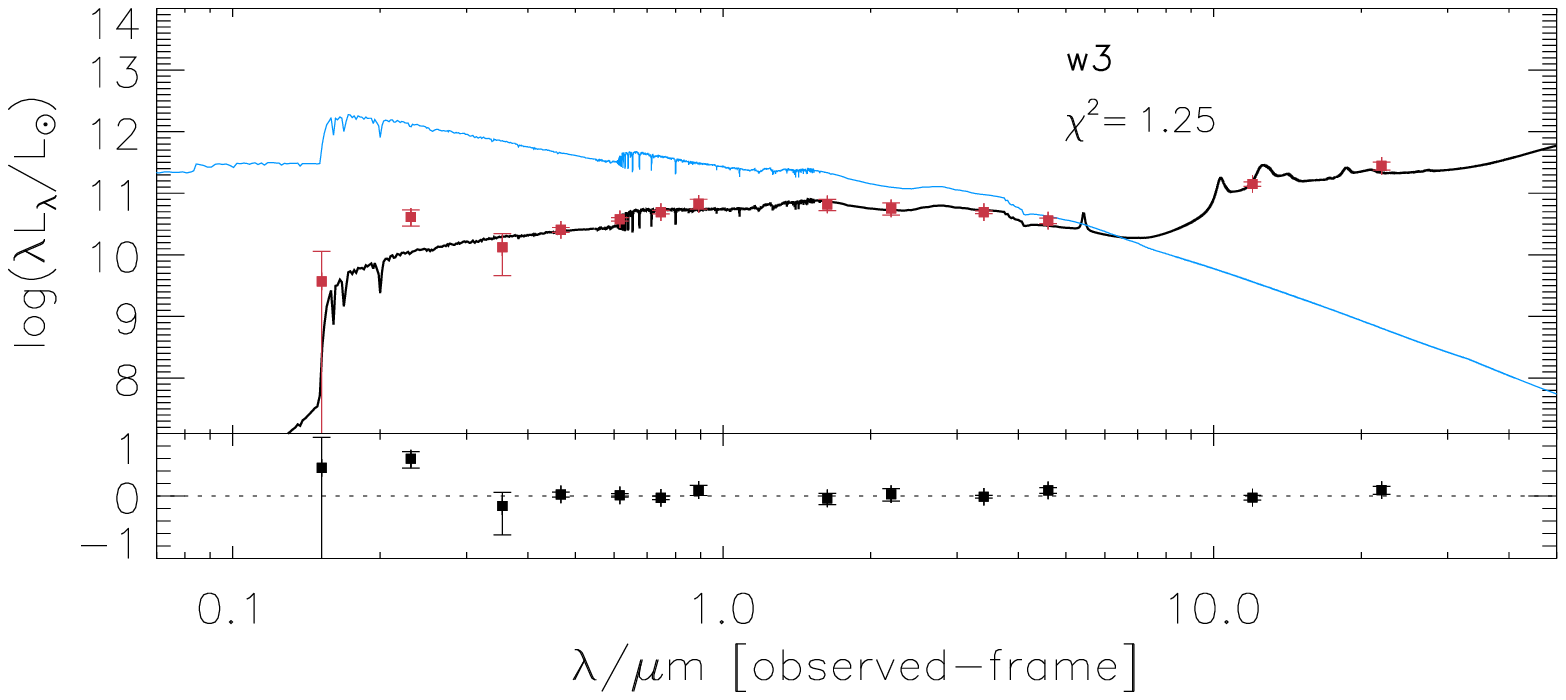} \\
\vspace{-1.5cm}
\includegraphics[scale=0.5,angle=0] {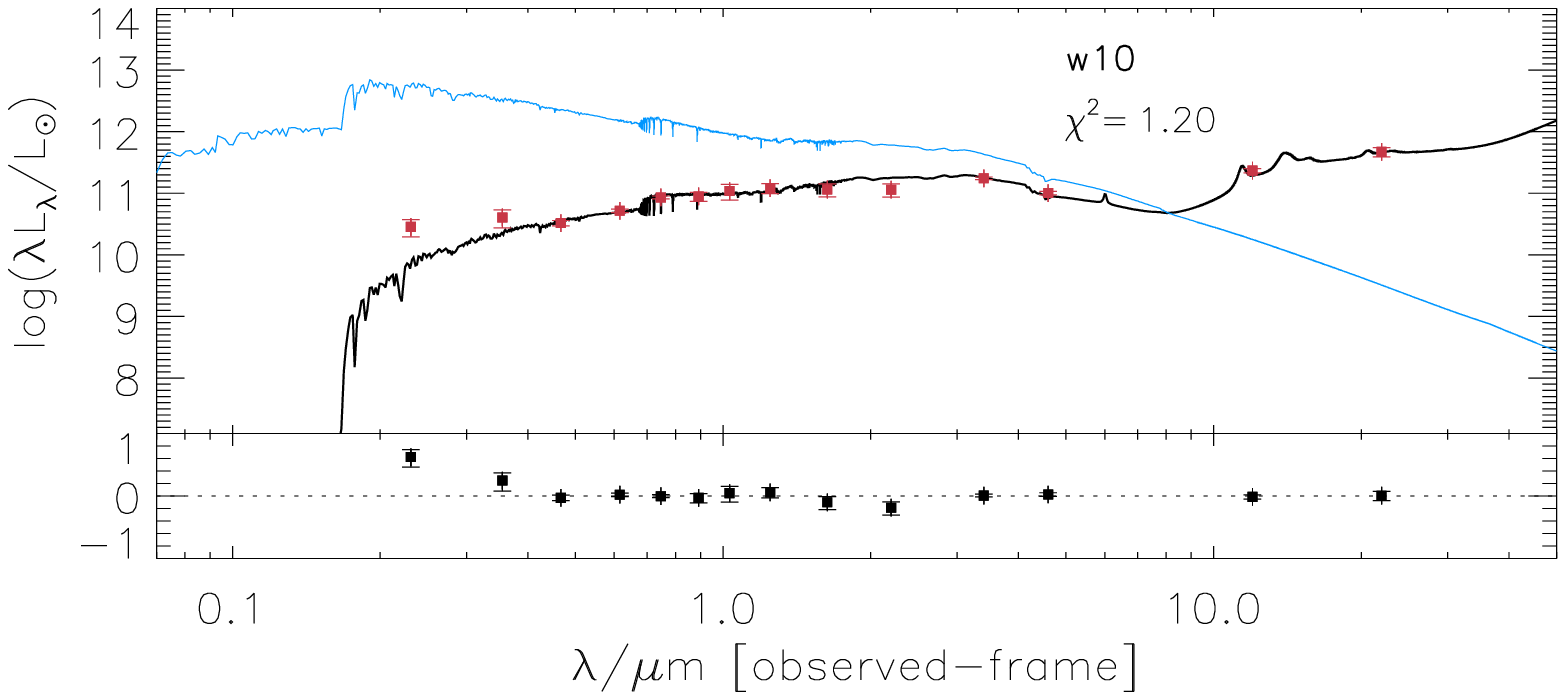} \\
\vspace{-1.5cm}
\includegraphics[scale=0.5,angle=0] {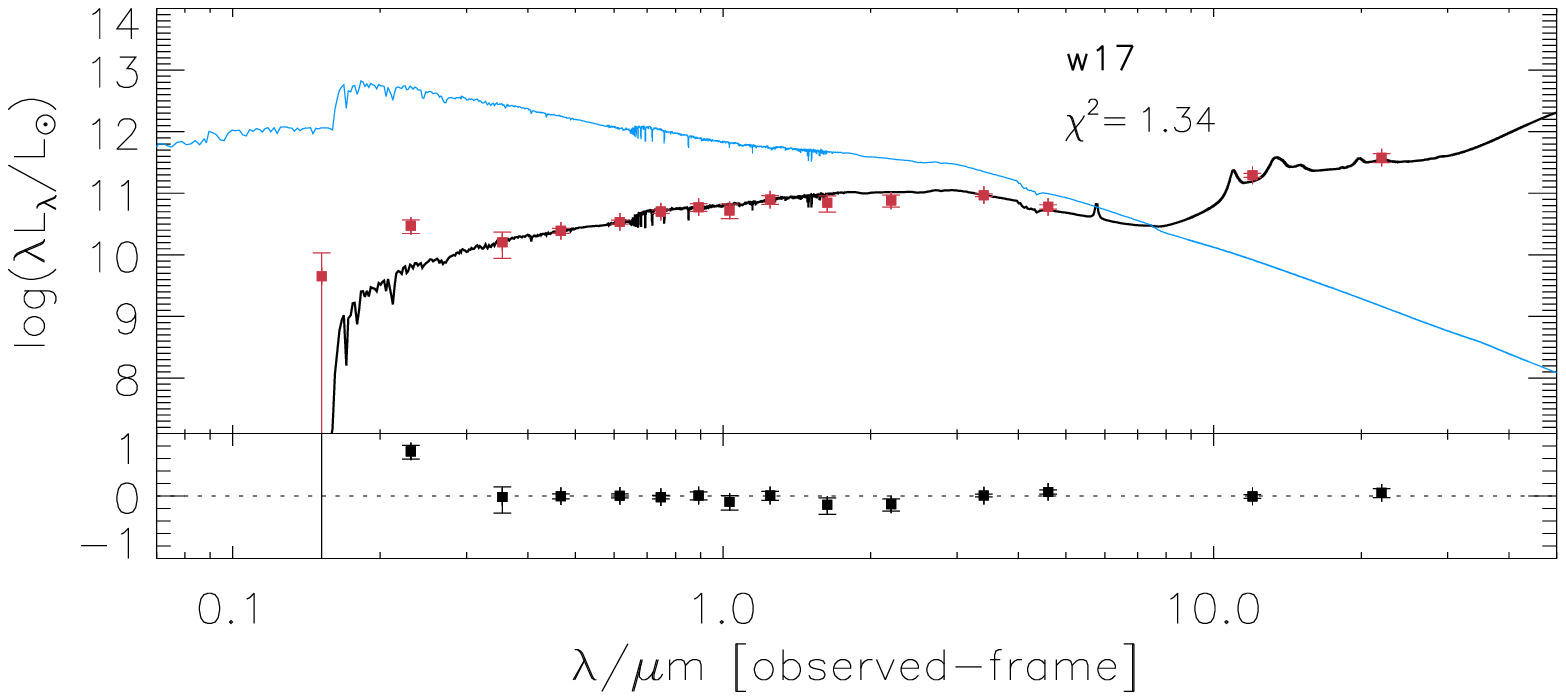} \\
\vspace{-1.5cm}
\includegraphics[scale=0.5,angle=0]{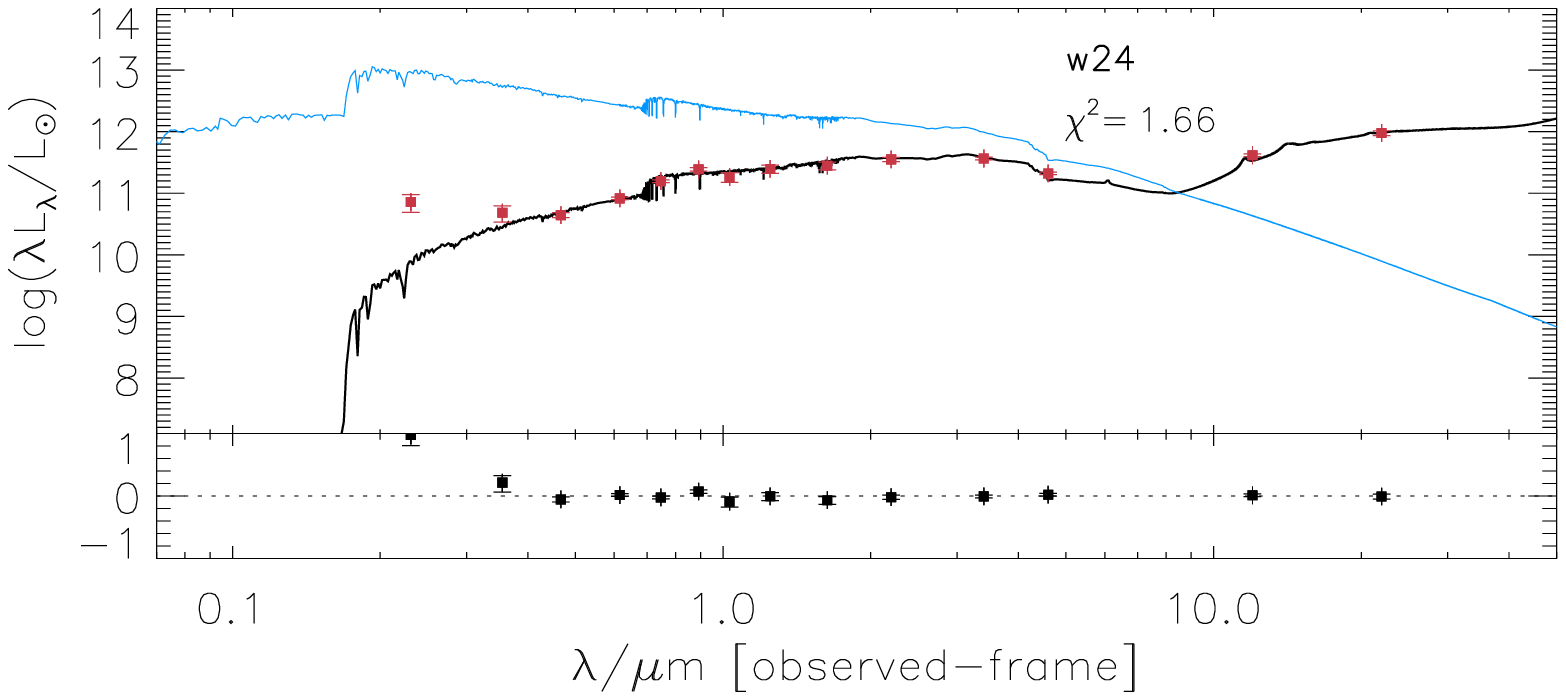} \\
\end{tabular}
\caption{Best-fit SEDs from MAGPHYS for some of the WiggleZ galaxies that are very luminous at 12 and 22$\mu$m. The plots show both the reddened and unreddened SEDs demonstrating that the galaxies have significant amounts of dust. Residuals from the reddened SED fits are plotted at the bottom of each panel.}
\label{fig:sed1}
\end{center}
\end{figure}

We fit the SEDs of the remaining 22 mid-IR bright star-formation dominated WiggleZ galaxies using the MAGPHYS code which provides a consistent treatment of the UV, optical and IR emission in star-forming galaxies. Some example SED fits can be seen in Figure \ref{fig:sed1} and encompass the range in properties seen in this small sample. In almost all cases, we find that the models under-predict the \textit{GALEX} UV fluxes of these galaxies and a UV excess is seen in the observed photometry relative to the models. This could be due to the effect of the Ly$\alpha$ line on the \textit{GALEX} fluxes of the galaxies or patchy dust extinction due to which some of the UV-light still remains unobscured. 

The median stellar mass of these mid-IR bright WiggleZ galaxies is log$_{10}$(M$_*$/M$_\odot$)=10.8$\pm$0.6 inferred from SED fits from the UV through to the mid IR, and they typically have star formation rates in excess of 100M$_\odot$yr$^{-1}$ and large dust extinction values. Their distribution in terms of redshift and observed UV-to-optical colours however, is the same as the rest of the WiggleZ sample. We conclude that a small fraction of the UV luminous WiggleZ galaxies at $z\sim$0.7 are also infrared luminous with SEDs consistent with young, dusty starbursts where some of the young stellar population is still unobscured in the UV. This is consistent with what is found for the $z\sim1$ Lyman Break Galaxy population some of which are also seen to be luminous in the \textit{Spitzer} MIPS 24$\mu$m band \citep{Burgarella:07}. In Appendix B, we provide the positions, redshifts and \textit{WISE} fluxes of these 22 mid infra-red luminous WiggleZ galaxies along with the best-fit stellar masses and star formation rates derived using MAGPHYS. 

\section{DISCUSSION}

Having constrained the stellar masses of a large sample of UV-luminous galaxies at $0.3<z<1.0$ and quantified the sensitivity of these mass estimates to assumptions made during the SED fitting process, we can now compare the stellar masses and SED fit parameters to other well-studied galaxy populations in order to place the WiggleZ galaxies within a global picture of galaxy evolution. Our study has concentrated on galaxies that populate the most extreme end of the blue-cloud and, as such, represent a sample where simple SED models are not likely to provide a good representation of the galaxy physics. Despite this, we have shown the stellar masses to be extremely robust to different assumptions made during the SED fitting with the median masses showing at most $\sim$0.3 dex variation due to differences in the SPS models and/or inclusion of additional bursts of star formation. Although the constraints on the star formation rates from the SED fitting are less robust with typical errors of $>$1 dex, the power of a large statistical sample such as ours is that the inferred \textit{median} properties of the galaxies can reasonably be taken to provide a good representation of the sample as a whole. 

The median star formation rate derived using single component star formation histories and FAST is between 3--10M$_\odot$yr$^{-1}$ regardless of the SPS model used and the number of photometric bands used in the SED fitting. The KG04 code gives a median star formation rate of $\sim$5 M$_\odot$yr$^{-1}$, which increases to only 5.3 M$_\odot$yr$^{-1}$ when secondary bursts are included. Finally, the subset of 1600 galaxies for which we compute best-fit SED parameters using MAGPHYS (Section \ref{sec:bursts}) also has a median star formation rate of 6--7M$_\odot$ yr$^{-1}$ consistent with the results from the other SED fitting codes. 

\begin{figure*}
\begin{center}
\begin{minipage}[c]{1.00\textwidth}
\centering
\includegraphics[width=8.5cm,height=6.0cm,angle=0] {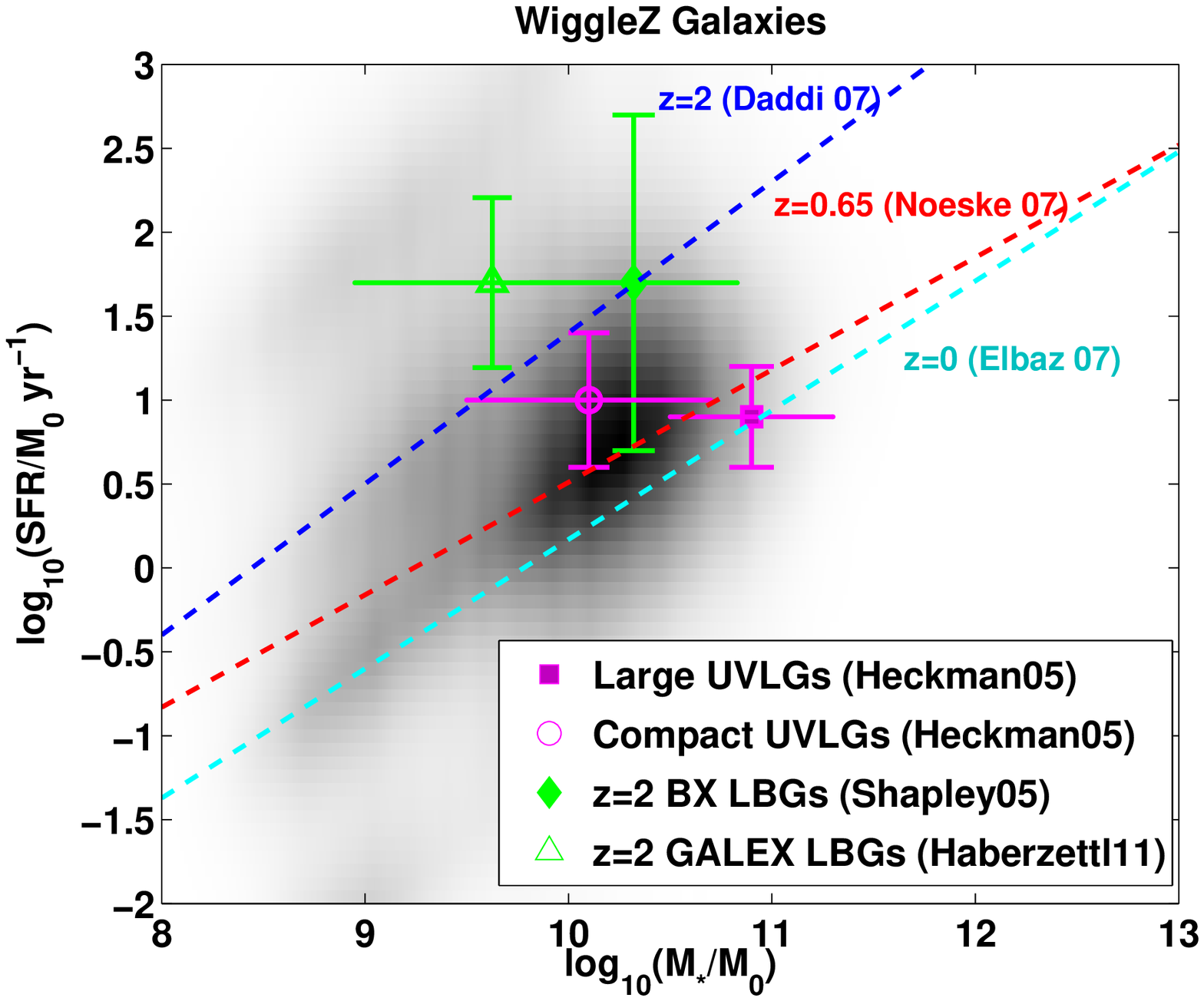}
\includegraphics[width=8.5cm,height=6.0cm,angle=0] {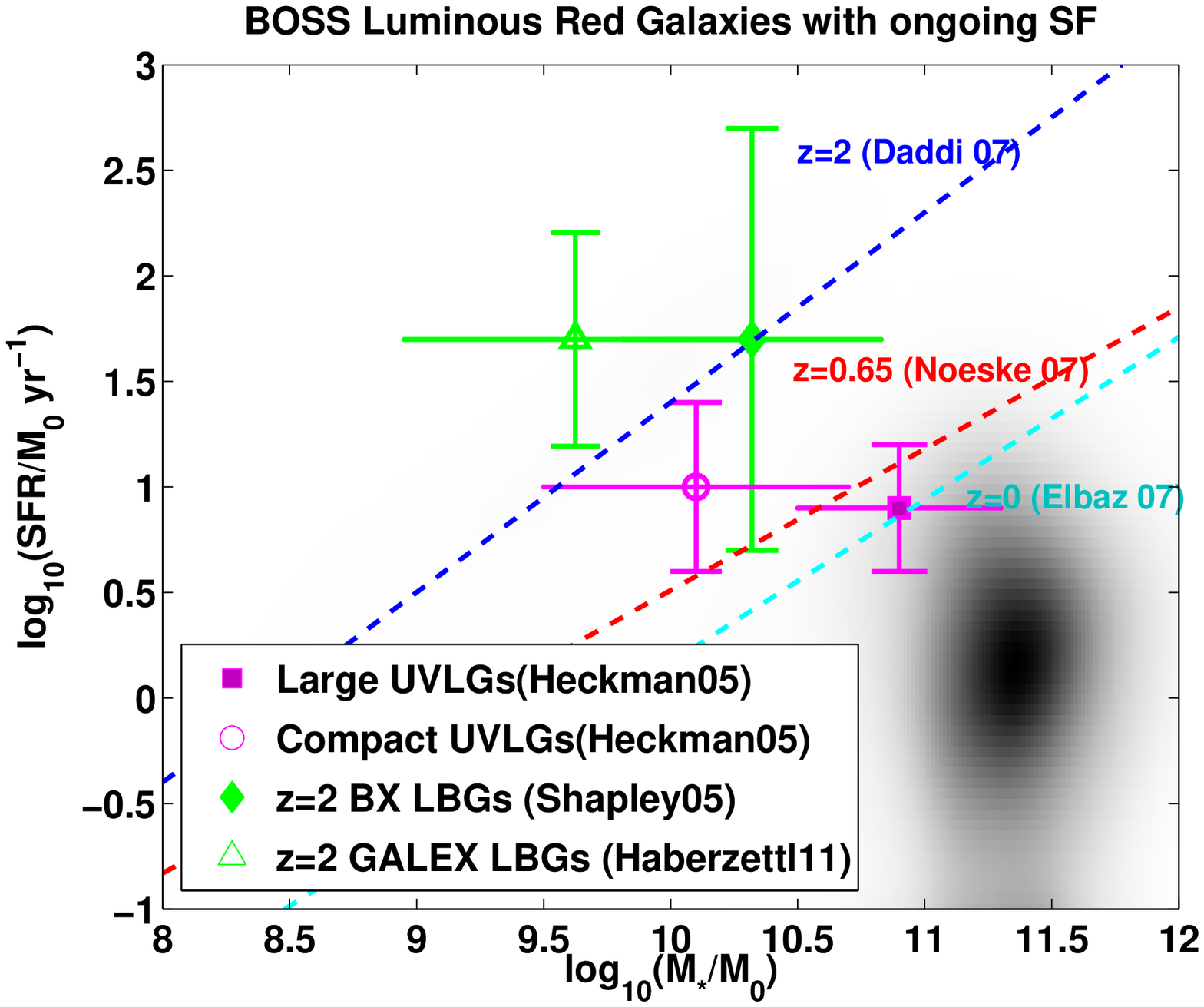}
\end{minipage}
\caption{Stellar mass versus star formation rate for all $\sim$40,000 WiggleZ galaxies (left) as well as Luminous Red Galaxies (LRGs) in the BOSS survey with ongoing star formation (right; \citealt{Maraston:12}). The greyscale represents the density of points in this plane. These distributions are compared to the average properties of local UV-luminous galaxies \citep{Heckman:05} as well as the $z\sim2$ Lyman break galaxy population \citep{Shapley:05, Haberzettl:11}. Dashed lines show the evolving main-sequence of star-forming galaxies at different redshifts \citep{Noeske:07, Elbaz:07, Daddi:07}. The WiggleZ galaxies mostly lie at the upper end of the main-sequence at $z\sim0.7$ with some clearly observed in a starburst phase. The LRGs on the other hand lie well below the main sequence.}
\label{fig:Mstar_SFR}
\end{center}
\end{figure*}

In Figure \ref{fig:Mstar_SFR} we plot the distribution of stellar masses and star formation rates for all $\sim$40,000 WiggleZ galaxies from this study. Spectroscopically confirmed Luminous Red Galaxies over the same redshift range represent a contrasting population of extremely red galaxies for comparison to the extremely blue WiggleZ galaxies.  Stellar masses have been estimated for large samples of these selected from the 2SLAQ \citep{Banerji:10} and SDSS-III-BOSS \citep{Maraston:12} surveys. For comparison to the WiggleZ population, we choose LRGs from BOSS that are fit to have ongoing star formation ($\sim$40 per cent of the full BOSS LRG sample from \citealt{Maraston:12}) and also show the location of these in the M$_*$--SFR plane in Figure \ref{fig:Mstar_SFR}. It is interesting to ask whether the very blue UV-luminous WiggleZ galaxies are the intermediate redshift analogues of the more distant Lyman Break and BX-selected galaxies at z$\sim$2 \citep{Adelberger:04, Shapley:05, Erb:06, Haberzettl:11} as well as local UV-luminous galaxies (UVLGs). \citet{Heckman:05} find that local UVLGs can be divided into large UVLGs and compact UVLGs. While the more massive large UVLGs have specific star formation rates (sSFR) sufficient to build their stellar mass over the Hubble time and therefore represent the most massive tail of star-forming disk galaxies like those in SDSS, the less massive compact UVLGs have higher sSFRs and are typically observed in a starbursting phase. The local UVLGs have typical $FUV$ luminosities of $\sim$3$\times10^{10}$L$_\odot$ while the higher redshift LBGs are slightly more luminous at $FUV$ wavelengths with typical FUV luminosities of $\sim$6$\times10^{10}$L$_\odot$. Our sample of WiggleZ galaxies has a median $FUV$ luminosity of $\sim$3$\times10^{10}$L$_\odot$ but at redshifts below 0.4, the median $FUV$ luminosity is only $\sim$9$\times10^{9}$L$_\odot$ whereas at $z>0.8$ it is typically $\sim$7$\times10^{10}$L$_\odot$. The low redshift WiggleZ galaxies are therefore less luminous than local UVLGs while the higher redshift WiggleZ galaxies have comparable FUV luminosities to LBGs. How do the rest of their properties compare to the UVLG and LBG samples? 

In Figure \ref{fig:Mstar_SFR}, we also plot the average stellar masses and SFRs of UVLGs taken from \citet{Heckman:05} and z$\sim$2 LBGs taken from \citet{Heckman:05, Shapley:05, Haberzettl:11} for comparison to the WiggleZ sample. These LBGs were selected either using GALEX \citep{Haberzettl:11} or the optical colour-selection: BX/BM technique \citep{Shapley:05}. The \textit{GALEX} selected population is less biased against star-forming galaxies with older stellar populations. Figure \ref{fig:Mstar_SFR} also shows the evolving `main-sequence' of star-forming galaxies taken from \citet{Elbaz:07} at $z=0$, \citet{Noeske:07} at $z=0.65$ roughly corresponding to the median redshift of our WiggleZ sample, and \citet{Daddi:07} at $z=2$. At higher redshifts, the main sequence systematically shifts to higher star formation rates for a given stellar mass, reflecting that high-redshift galaxies were more active than those in the local Universe. While it is difficult to empirically constrain any M$_*$--SFR relation from our WiggleZ sample, which is highly incomplete in many regions of this parameter space, we can nevertheless ask where the WiggleZ galaxies lie relative to the already derived M$_*$--SFR relations from previous works. We have already noted that large UVLGs lie on the main sequence at $z=0$ while the compact UVLGs show excess star formation relative to this main sequence. Similarly, the BX/BM LBGs also lie on the $z=2$ main sequence and therefore represent normal star-forming galaxies at these redshifts, while the \textit{GALEX} selected LBGs show excess star formation relative to the $z=2$ main sequence. Figure \ref{fig:Mstar_SFR} shows that the WiggleZ galaxies represent a heterogenous population with most lying at the upper end of the $z=0.65$ main sequence, but also a cloud of galaxies above this main sequence, that are typically observed in a bursting phase (see also Jurek et al., in preparation). Although the scatter seen in Figure \ref{fig:Mstar_SFR} is dominated by the scatter in the determination of the SFR from the SED fitting, there is evidence from those galaxies where the SFR is relatively well constrained, that WiggleZ galaxies with bluer optical $(g-r)$ colours predominantly lie above the main sequence. At a redshift of $z=0.7$, a typical galaxy of 10$^{10}$M$_\odot$ would need to be forming stars at a rate of $\sim$1.4M$_\odot$yr$^{-1}$ in order to build up it's entire stellar mass through constant star formation over the age of the Universe.  Most of the WiggleZ galaxies have star formation rates that are higher than this.

We also note that the stellar masses of the WiggleZ galaxies are comparable to the local \textit{GALEX} selected UVLGs, as well as the well-studied high-redshift LBG population. In Section \ref{sec:bursts}, we found that the WiggleZ galaxies show a trend of increasing dust attenuation with increasing stellar mass and decreasing $FUV$ luminosity as also observed in higher redshift UV-luminous galaxies. These WiggleZ galaxies may therefore reasonably be thought to represent the intermediate redshift analogues of the local UVLG and LBG populations.   

We note by contrast that the star-forming LRGs at similar redshifts lie well away from the main-sequence and have star formation rates that are more than an order of magnitude lower than the main-sequence of star-forming galaxies. This is consistent with these massive red galaxies having undergone the bulk of their stellar mass assembly at earlier epochs, when they presumably also experienced much higher levels of star formation. 

\section{CONCLUSION}

We have conducted a detailed study of the stellar masses of a large sample of $\sim$40,000 UV-luminous spectroscopically confirmed galaxies at 0.3$<$z$<$1.0 selected within the WiggleZ survey. Around 30 per cent of the sample are matched to the wide-field NIR UKIDSS Large Area Survey and around 15 per cent are additionally detected at 3.4 and 4.6$\mu$m in the all-sky \textit{WISE} survey. The IR detected population represents the redder, more luminous and more massive end of the WiggleZ population with stellar masses that are on average 0.6 dex larger for the UKIDSS detected galaxies and 0.8 dex larger for the \textit{WISE} detected galaxies. However, at the high redshift end of the WiggleZ sample at $z > 0.7$, there is evidence for a cloud of IR undetected galaxies that are just as massive and just as red in terms of their optical colours, as the IR detected galaxies. In addition, we find a small sample of 22 galaxies which are also extremely luminous at 12 and 22$\mu$m with SEDs consistent with dusty starburst galaxies where some of the younger stellar population is still unobscured in the UV. 

As the WiggleZ galaxies represent the most extreme end of the blue cloud population at these redshifts, where SED fitting is likely to be the most problematic, we quantify the sensitivity of our stellar mass estimates to assumptions made during the SED fitting process. In particular we find that:

\begin{itemize}

\item{The effect that the NIR photometry has in constraining the stellar masses, depends on the priors assumed in the SED fitting. With the SED fitting code, FAST used in conjunction with the BC03 and Maraston models, we find that the stellar mass constraints are improved on addition of the NIR photometry as this allows us to tighten the upper bound on the stellar mass estimates. With the KG04 code and the PEGASE.2 SPS models, we find that the NIR data makes little difference to the stellar masses and their corresponding errors. Regardless of the choice of SED fitting code, the optical and optical+NIR derived masses are consistent within the 1$\sigma$ errors for $>$68\% of the galaxies as expected. With the inclusion of NIR photometry, the mass estimates from both FAST and KG04 agree very well.}

\item{The addition of UV photometry from $GALEX$ makes very little difference to both the median stellar masses and the median 1$\sigma$ errors on these stellar masses, even for these extremely blue WiggleZ galaxies selected in the UV. However, there is a small population of IR-detected WiggleZ galaxies where the lack of UV photometry leads to best-fit SEDs with considerably higher star formation rates, dust extinctions and younger ages than when the UV photometry is included in the fitting. A consequence of fitting younger stellar populations without the UV data, is that the stellar masses of these galaxies are also under-estimated without the UV.}

\item{The choice of SPS model can affect the stellar mass estimates by $\sim$0.3 dex. The Maraston models which include TP-AGB stars result in stellar masses that are $\sim$0.1--0.2 dex lower than those inferred using the BC03 models with the differences largest for the more $FUV$ luminous subset of WiggleZ galaxies, which are typically older. The PEGASE.2 models produce stellar masses that are 0.2(0.3) dex higher than BC03 (Maraston). The effect is more pronounced for galaxies with fainter $FUV$ luminosities.}

\item{The inclusion of nebular emission in the SED models, lowers the stellar masses by only $\sim$0.02 dex on average. This offset is smaller than that seen in the case of high redshift LBGs.}


\item{The choice of IMF can affect the stellar mass by up to 0.3 dex. The Salpeter IMF produces the largest stellar masses. The Kroupa IMF results in stellar masses that are $\sim$0.1--0.2 dex lower than Salpeter depending on the SPS model being fit. The BG03 and Chabrier IMFs produce very similar stellar masses with a median difference of $<$0.001 dex between them. These masses are $\sim$0.24 dex lower than those derived using the Salpeter IMF and $\sim$0.1 dex lower than those derived using the Kroupa IMF.}

\item{For single component SFHs, the stellar masses are insensitive to the choice of star formation history although we find as expected that the truncated SFH does not provide good $\chi^2$ values for these actively star-forming galaxies. The addition of bursts of star formation on top of the smooth underlying SFH however, results in stellar masses that are up to 0.3 dex larger. The effect is more pronounced at higher $FUV$ luminosities where secondary bursts can effectively hide an older, more massive and more evolved stellar population. When bursts are allowed, the typical burst fraction is estimated to be $\sim$1 per cent of the total stellar mass in these galaxies.}

\item{Although the dust extinction parameter, A$_V$ is not as well constrained as the stellar mass, we note a trend of increasing dust attenuation with increasing stellar mass and decreasing $FUV$ luminosity in these galaxies. These trends are consistent with what has been found in higher redshift UV-luminous galaxies.}

\item{We compare our best-fit rest-frame $K$-band mass-to-light ratios from the SED fits to the predictions from simple optical colour based estimators such as those of \citet{Bell:03}. We find that the colour-based estimator over-predicts the mass-to-light ratio for $\sim$75\% of the WiggleZ galaxies where the results from this estimator and the SED fits are formally inconsistent given the 1$\sigma$ errors. The median difference in M/L$_K$ between the simple colour-based estimator, and the more sophisticated SED fitting approach, is $\sim$0.4 dex. The inconsistencies are most pronounced for the bluer WiggleZ galaxies with young best-fit ages and galaxies with ages $\gtrsim$7Gyr generally have M/L$_K$ consistent with the colour-based estimator.}

\end{itemize}

We conclude that our stellar mass estimates have typical dispersions of $\sim$0.1-0.3 dex as a result of changes in the input parameters of the SED fitting and they are therefore extremely robust to these changes even for these extremely blue galaxies. While the star formation rates have much poorer constraints from the photometric data, the advantage of a large sample such as ours is that the \textit{median} values inferred from the SED fits, should provide a reasonable representation of the average properties of the WiggleZ sample as a whole. We find that the WiggleZ galaxies have star formation rates of $\sim$3--10M$_\odot$yr$^{-1}$ and these estimates are consistent between the different SED fitting codes tested in this work. Although these star formation rate estimates have very large errors for individual galaxies, we compare the distribution of the WiggleZ galaxies in the M$_*$--SFR plane to other well-studied samples such as Luminous Red Galaxies over the same redshift range, local UV-luminous galaxies, and more distant Lyman Break Galaxies. We find that the WiggleZ population, on average lies at the upper end of the main-sequence of star-forming galaxies at $z\sim0.7$. However, there is evidence for some of the bluer galaxies lying well above the main sequence and therefore being observed in a starburst phase (Jurek et al., in preparation). The stellar masses of our WiggleZ galaxies are comparable to both $z\sim2$ LBGs and compact UV-luminous galaxies in the local Universe suggesting that this population can be taken to represent a reasonable intermediate redshift analogue that straddles the redshift space between the well studied local UVLG and distant LBG populations.  

We conclude that the combination of current cosmological volume spectroscopic surveys and wide-field photometric surveys, provides a promising route for constraining the physical properties of very large, statistically robust samples of intermediate redshift galaxies. Understanding the properties of these z$\sim$1 galaxies is essential for bridging the gap between the local and high redshift Universe, as well as devising effective target selection criteria for the next generation of wide-field spectroscopic surveys.  

The stellar masses in this paper will be made publicly available with the next WiggleZ data release.

\label{sec:conclusions}

\section*{Acknowledgements}

We thank the anonymous referee for a constructive report that has helped improve this paper. MB acknowledges Paul Hewett and Richard McMahon for many constructive discussions and Claudia Maraston and Joel Brownstein for access to the BOSS LRG stellar masses. MB wishes to acknowedge financial support from the STFC through grants held both at the Institute of Astronomy, Cambridge and University College London. The authors acknowledge financial support from The Australian Research Council (grants DP1093738, DP0772084. LX0881951, LE0668442), Swinburne University of Technology, The University of Queensland, and the Anglo-Australian Observatory for the WiggleZ survey. The WiggleZ survey would not be possible without the dedicated work of the staff of the Australian Astronomical Observatory in the development and support of the AAOmega spectrograph, and the running of the AAT. K.G. also acknowledges support from Australian Research Council grant DP1094370 for galaxy evolution studies.

\textit{GALEX} (the Galaxy Evolution Explorer) is a NASA Small Explorer, launched in April 2003. We gratefully ac- knowledge NASAs support for construction, operation and science analysis for the GALEX mission, developed in co- operation with the Centre National dEtudes Spatiales of France and the Korean Ministry of Science and Technology.


\bibliography{}

\begin{appendix}

\section{Example SED Fits}

For illustrative purposes, we present some examples of the SED fits for the WiggleZ galaxies derived using the KG04 code with multi-component SFHs and the PEGASE.2 SPS models including nebular emission. These example SEDs have specifically been chosen to encompass the full range in properties of our sample in terms of redshift, colour, stellar mass and age and also include some examples of poorly fit SEDs.  

\begin{figure*}
\begin{center}
\begin{tabular}{c}
\includegraphics[scale=0.35]{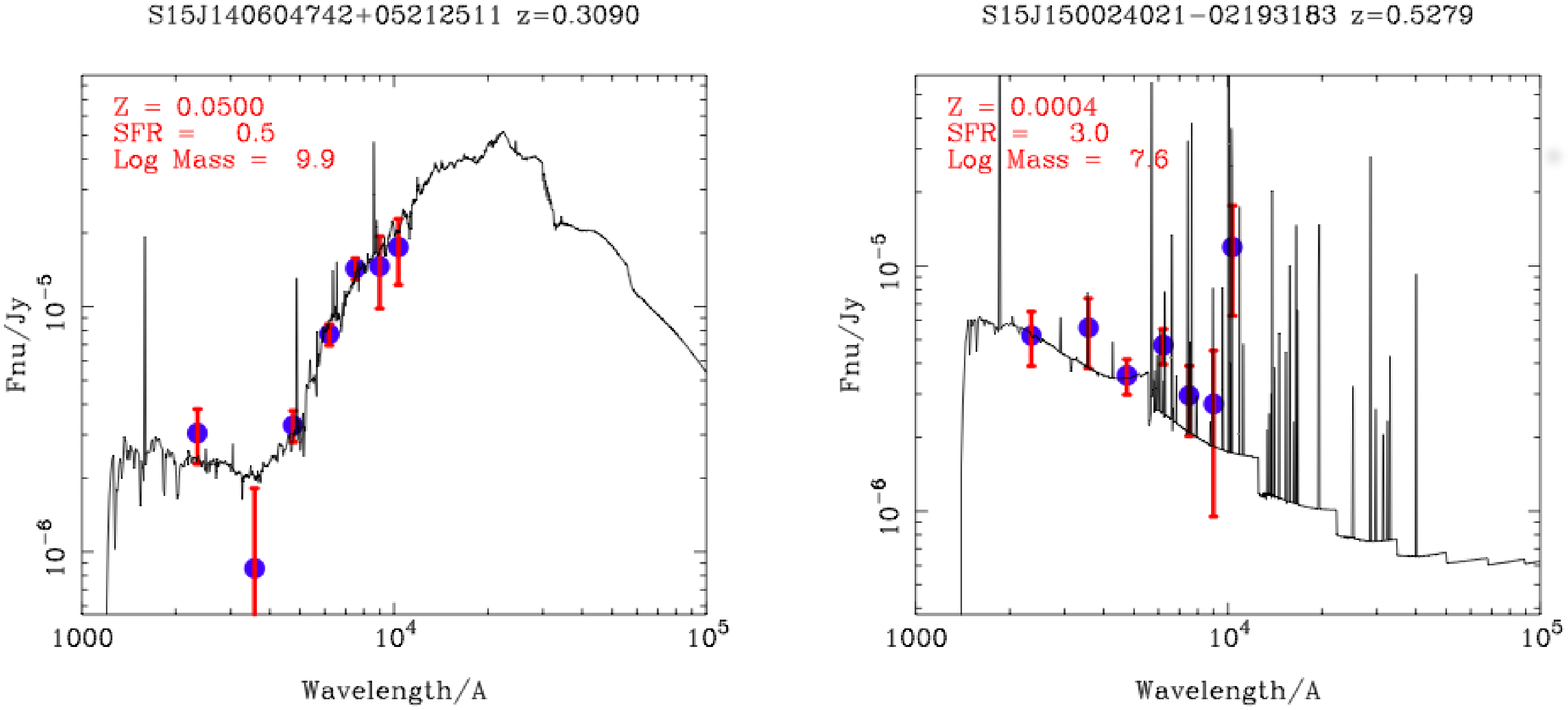} \\
\includegraphics[scale=0.35]{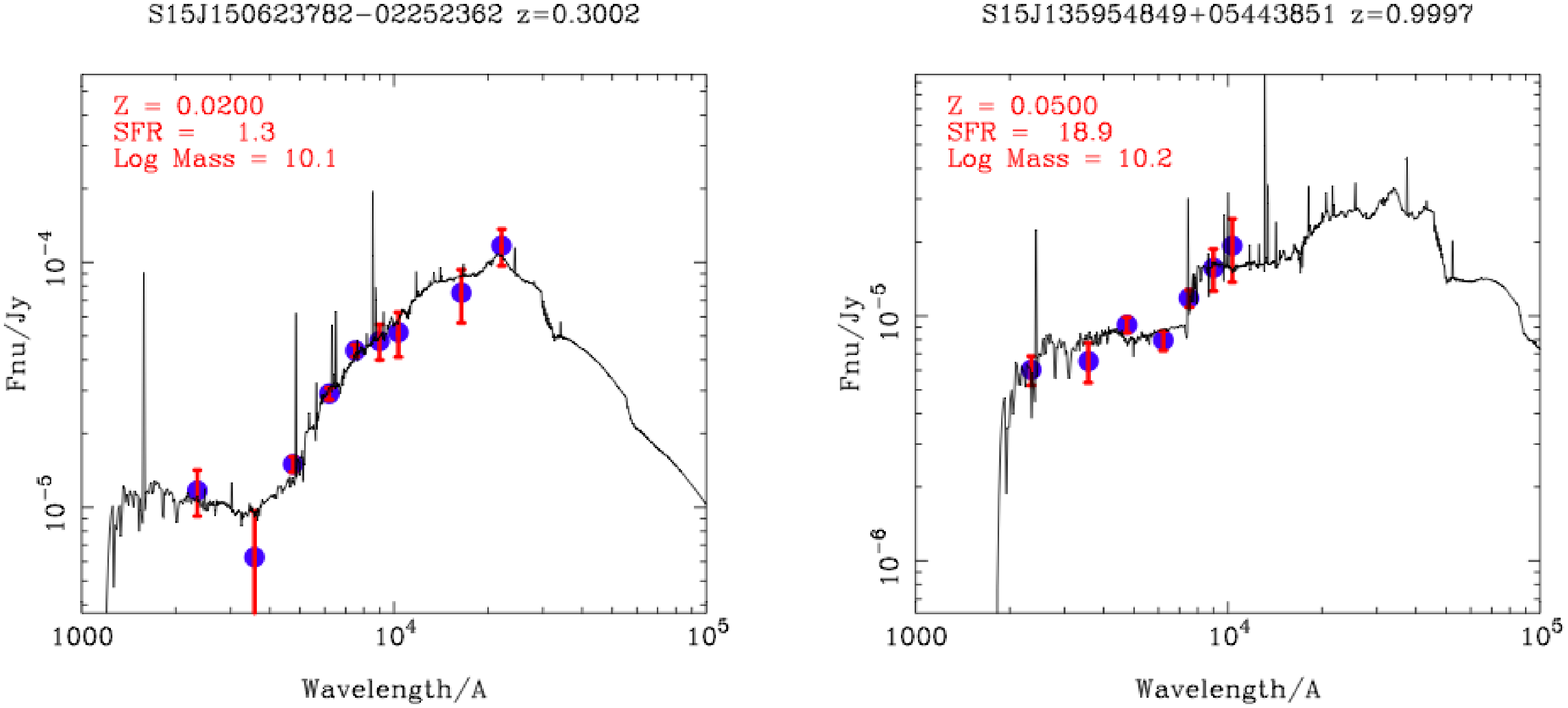} \\
\includegraphics[scale=0.35]{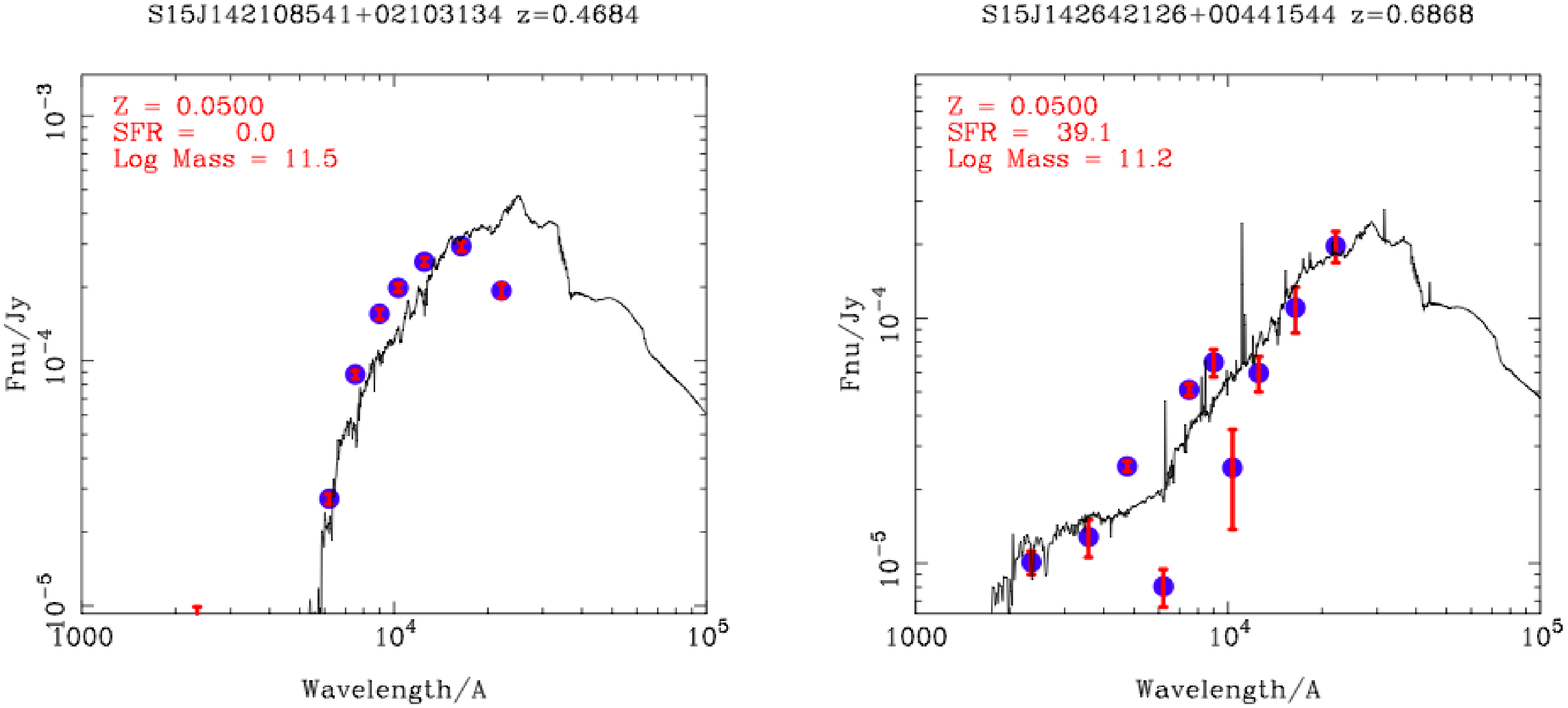} \\
\includegraphics[scale=0.35]{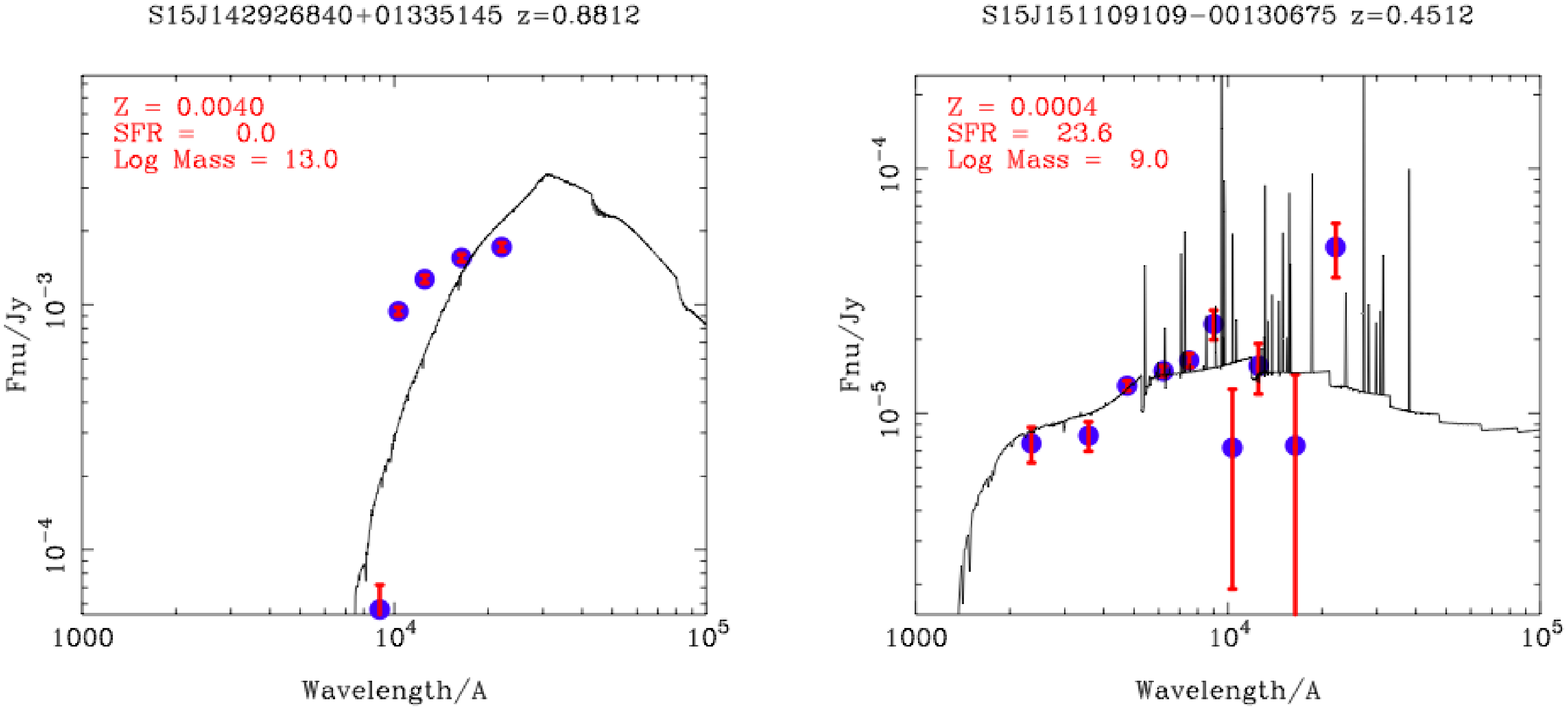} \\
\end{tabular}
\caption{Example SED fits encompassing the full range of properties in our sample and showing some examples of galaxies that are poorly fit as well. These fits are derived using the KG04 code and the PEGASE.2 SPS models and including nebular emission lines.}
\end{center}
\end{figure*}

\section{Mid-infrared luminous WiggleZ galaxies}

We presented a small sample of mid infrared luminous WiggleZ galaxies detected at 12 and 22$\mu$m in \textit{WISE}. Based on the \textit{WISE} colours, these galaxies do not seem to have a significant AGN component and can be considered to be star-formation dominated. Here we provide a table of the positions, redshifts and \textit{WISE} fluxes for these 22 galaxies in Table B1. The best-fit stellar mass and star formation rate derived using MAGPHYS are also given in the same Table.  

\onecolumn
\begin{landscape}
\begin{table*}
\label{tab:24}
\caption{Summary of \textit{WISE} fluxes, best-fit stellar mass and star formation rates derived using MAGPHYS for the 22 mid infrared luminous WiggleZ galaxies detected at 12 and 22$\mu$m in \textit{WISE}, and where the colours rule out a significant AGN component.}
\begin{center}
\begin{tabular}{cccccccccc}
\hline
Name & Redshift & RA & DEC & 3.4$\mu$m (mJy) & 4.6$\mu$m (mJy) & 12$\mu$m (mJy) & 22$\mu$m (mJy) & log$_{10}$(M$_*$/M$_\odot$) & SFR(M$_\odot$yr$^{-1}$) \\
\hline \hline
w2 & 0.532 & 14:34:53.298 &$-$02:29:10.75 & 0.219$\pm$0.008 & 0.217$\pm$0.012 & 1.044$\pm$0.091 & 6.415$\pm$0.739 & 10.8$^{+0.4}_{-0.1}$ & 120$^{+230}_{-90}$ \\
w3 & 0.647 & 14:10:26.451 & $-$02:04:01.30 & 0.120$\pm$0.006 & 0.118$\pm$0.012 & 1.208$\pm$0.105 & 4.365$\pm$0.631 & 10.4$^{+0.4}_{-0.2}$ & 150$^{+170}_{-90}$ \\
w4 & 0.681 & 14:28:03.843 & $-$01:15:08.36 & 0.250$\pm$0.009 & 0.186$\pm$0.013 & 1.934$\pm$0.103 & 4.349$\pm$0.745 & 10.5$^{+0.0}_{-0.0}$ & 220$^{+5}_{-10}$ \\
w5 & 0.428 & 15:06:56.064 & $+$01:13:17.75 & 0.232$\pm$0.008 & 0.243$\pm$0.013 & 1.271$\pm$0.091 & 4.716$\pm$0.725 & 11.1$^{+0.1}_{-0.4}$ & 130$^{+120}_{-60}$ \\
w6 & 0.876 & 14:58:28.703 & $+$01:18:48.49 & 0.194$\pm$0.007 & 0.188$\pm$0.011 & 0.923$\pm$0.084 & 3.688$\pm$0.642 & 10.5$^{+0.4}_{-0.0}$ & 210$^{+2}_{-20}$ \\
w7 & 0.719 & 14:22:16.253 & $+$01:23:38.10 & 0.192$\pm$0.008 & 0.161$\pm$0.012 & 1.181$\pm$0.091 & 3.644$\pm$0.654 & 11.0$^{+0.0}_{-0.2}$ & 1200$^{+20}_{-1100}$ \\
w8 & 0.446 & 14:25:03.516 & $+$01:37:20.61 & 0.234$\pm$0.008 & 0.246$\pm$0.013 & 1.427$\pm$0.092 & 8.002$\pm$0.752 & 11.1$^{+0.0}_{-0.1}$ & 190$^{+0.0}_{-10}$ \\
w9 & 0.738 & 14:41:34.600 & $+$02:30:08.30 & 0.240$\pm$0.009 & 0.157$\pm$0.011 & 0.573$\pm$0.083 & 4.153$\pm$0.673 & 10.7$^{+0.1}_{-0.1}$ & 270$^{+80}_{-70}$ \\
w10 & 0.821 & 14:40:47.785 & $+$03:26:21.70 &  0.236$\pm$0.008 & 0.184$\pm$0.011 & 1.109$\pm$0.090 & 4.096$\pm$0.694 & 10.9$^{+0.4}_{-0.1}$ & 700$^{+80}_{-570}$ \\
w11 & 1.102 & 15:11:46.032 & $+$03:28:52.61 & 0.197$\pm$0.008 & 0.191$\pm$0.011 & 1.025$\pm$0.085 & 4.254$\pm$0.682 & 11.3$^{+0.0}_{-0.0}$ & 1950$^{+30}_{-30}$ \\
w12 & 0.503 & 14:33:32.897 & $+$03:36:10.61 &  0.130$\pm$0.006 & 0.109$\pm$0.010 & 0.869$\pm$0.086 & 4.414$\pm$0.772 & 10.7$^{+0.2}_{-0.1}$ & 14$^{+30}_{-3}$ \\
w13 & 0.771 & 14:49:30.259 & $+$03:36:26.17 & 0.178$\pm$0.007 & 0.187$\pm$0.013 & 1.576$\pm$0.094 & 4.890$\pm$0.761 & 11.2$^{+0.1}_{-0.2}$ & 320$^{+30}_{-210}$ \\
w14 & 0.490 & 14:07:12.577 & $+$03:37:05.23 &  0.100$\pm$0.006 & 0.097$\pm$0.011 & 0.688$\pm$0.091 & 5.345$\pm$0.684 & 10.1$^{+0.4}_{-0.1}$ & 100$^{+20}_{-90}$ \\
w15 & 0.602 & 14:55:33.146 & $+$03:37:17.67 & 0.209$\pm$0.008 & 0.147$\pm$0.011 & 0.970$\pm$0.086 & 4.089$\pm$0.678 & 10.5$^{+0.0}_{-0.1}$ & 140$^{+70}_{-20}$ \\
w16 & 0.530 & 14:35:04.809 & $+$03:50:16.73 &  0.221$\pm$0.008 & 0.190$\pm$0.012 & 1.102$\pm$0.089 & 6.748$\pm$0.684 & 11.0$^{+0.0}_{-0.3}$ & 160$^{+190}_{-60}$ \\
w17 & 0.749 & 14:53:38.175 & $+$04:34:20.93 & 0.157$\pm$0.007 & 0.138$\pm$0.011 & 1.167$\pm$0.085 & 4.131$\pm$0.677 & 10.9$^{+0.4}_{-0.3}$ & 450$^{+50}_{-400}$ \\
w18 & 0.507 & 15:11:58.404 & $+$04:56:23.95 & 0.241$\pm$0.008 & 0.248$\pm$0.012 & 1.608$\pm$0.086 & 3.506$\pm$0.594 & 10.8$^{+0.0}_{-0.0}$ & 0.02$^{+0.001}_{-0.001}$ \\
w19 & 0.670 & 14:40:37.878 & $+$05:06:37.37 & 0.164$\pm$0.007 & 0.165$\pm$0.011 & 0.942$\pm$0.088 & 4.357$\pm$0.670 & 10.6$^{+0.8}_{-0.0}$ & 440$^{+4}_{-320}$ \\
w21 & 0.726 & 14:21:01.776 & $+$05:22:17.11 & 0.154$\pm$0.006 & 0.123$\pm$0.011 & 0.700$\pm$0.083 & 2.935$\pm$0.560 & 10.3$^{+0.3}_{-0.0}$ & 140$^{+130}_{-40}$ \\
w22 & 0.827 & 14:06:00.312 & $+$05:49:07.65 & 0.075$\pm$0.005 & 0.068$\pm$0.011 & 0.760$\pm$0.087 & 3.869$\pm$0.663 & 10.8$^{+0.2}_{-0.2}$ & 160$^{+3}_{-120}$ \\
w23  & 0.559 & 14:09:51.618 & $+$05:51:04.54 &  0.160$\pm$0.007 & 0.161$\pm$0.011 & 0.668$\pm$0.084 & 4.521$\pm$0.675 & 11.2$^{+0.0}_{-0.6}$ & 90$^{+0}_{-70}$\\
w24 & 0.844 & 14:59:15.761 & $+$06:17:19.93 & 0.462$\pm$0.012 & 0.358$\pm$0.014 & 1.842$\pm$0.092 & 7.755$\pm$0.707 & 11.4$^{+0.1}_{-0.1}$ & 1050$^{+540}_{-440}$\\
\hline
\end{tabular}
\end{center}
\end{table*}
\end{landscape}
\twocolumn

\end{appendix}

\end{document}